\documentclass[aps,prb,twocolumn,superscriptaddress,showpacs]{revtex4-1}

\usepackage{amsmath}
\usepackage{amssymb}
\usepackage{graphicx}
\usepackage{epsfig}
\usepackage{multirow}
\usepackage{amsmath}
\usepackage{amssymb}

\newcommand{\ep}{\varepsilon}

\newcommand{\bs}[1]{\boldsymbol{#1}}

\begin{document}


\title{Collision-dominated nonlinear hydrodynamics in graphene}


\author{U. Briskot}
\affiliation{Institute of Nanotechnology, Karlsruhe Institute of Technology, 76021 Karlsruhe, Germany}
\affiliation{Institute for Theoretical Condensed Matter Physics, Karlsruhe Institute of Technology, 76128 Karlsruhe, Germany}

\author{M. Sch\"utt}
\affiliation{School of Physics and Astronomy, University of Minnesota, Minneapolis 55455, USA}

\author{I.V. Gornyi}
\affiliation{Institute of Nanotechnology, Karlsruhe Institute of Technology, 76021 Karlsruhe, Germany}
\affiliation{Institute for Theoretical Condensed Matter Physics, Karlsruhe Institute of Technology, 76128 Karlsruhe, Germany}
\affiliation{Ioffe Physical Technical Institute, 194021 St. Petersburg, Russia}

\author {M. Titov}
\affiliation{Radboud University Nijmegen, Institute for Molecules and Materials, NL-6525 AJ Nijmegen, The Netherlands}

\author{B.N. Narozhny}
\affiliation{Institute for Theoretical Condensed Matter Physics, Karlsruhe Institute of Technology, 76128 Karlsruhe, Germany}
\affiliation{National Research Nuclear University MEPhI (Moscow Engineering Physics Institute), 115409 Moscow, Russia}

\author{A.D. Mirlin}
\affiliation{Institute of Nanotechnology, Karlsruhe Institute of Technology, 76021 Karlsruhe, Germany}
\affiliation{Institute for Theoretical Condensed Matter Physics, Karlsruhe Institute of Technology, 76128 Karlsruhe, Germany}
\affiliation{Petersburg Nuclear Physics Institute, 188300 St. Petersburg, Russia}

\date{\today}

\begin{abstract}
 We present an effective hydrodynamic theory of electronic transport
 in graphene in the interaction-dominated regime. We derive the
 emergent hydrodynamic description from the microscopic Boltzmann
 kinetic equation taking into account dissipation due to Coulomb
 interaction and find the viscosity of Dirac fermions in graphene for
 arbitrary densities. The viscous terms have a dramatic effect on
 transport coefficients in clean samples at high temperatures. Within
 linear response, we show that viscosity manifests itself in the
 nonlocal conductivity as well as dispersion of hydrodynamic plasmons.
 Beyond linear response, we apply the derived nonlinear hydrodynamics
 to the problem of hot spot relaxation in graphene.
\end{abstract}


\maketitle

Physics at long time and length scales can be conveniently described
within the hydrodynamic approach \cite{Landau6}. The appeal of this
approach is hinged on its ability to describe a wide range of physical
systems \cite{Landau10,wolf} using the same, relatively small set of
quantities and equations governing their behavior. At the same time,
the final form of the hydrodynamic equations varies from system to
system \cite{Landau6,wolf} reflecting the particular symmetries and
other physical features of the problem.

Traditional hydrodynamics \cite{Landau6} describes the system in terms
of the velocity field $\bs{v}$. The equations describing the velocity
field (e.g., the Euler equation in the case of the ideal liquid or the
Navier-Stokes equation if dissipation is taken into account) can be
either inferred from symmetry arguments or derived from the Boltzmann
kinetic equation.  Both approached require one to express the fluxes
of conserved quantities (energy, momentum, etc.) in terms of $\bs{v}$.
In particular, the viscous terms appearing in the Navier-Stokes
equation can be traced to a particular approximation for the momentum
flux (or the stress tensor) $\Pi_{\alpha\beta}$. The specific form of
$\Pi_{\alpha\beta}$ depends on whether one discusses a usual,
Galilean-invariant or a relativistic, Lorentz-invariant system.

Low-energy excitations in graphene \cite{Novoselov2004} present a most
interesting case of a system that is neither Galilean- nor
Lorentz-invariant. This poses a significant challenge in establishing
the hydrodynamic description in graphene, which has to be derived from
first principles
\cite{hart,FritzSchmalianCritTransp2008,mu11,MuellerSachdev2008,mu12,FosterAleiner2009,SchmalianViscosity2009,Svintsov2012,tom2013,Svintsov2013,levitov,levitov2013,tom2014,LinHydroBoris2015,prin}. The
resulting equations should account for physical processes at time and
length scales that are much longer than the scales related to the
microscopic processes responsible for equilibration of the system.
The issue of scale separation is especially important in the vicinity
of charge neutrality in clean graphene.  Without interaction there is
no ``intrinsic'' energy scale other than temperature.

The interest in hydrodynamics in graphene has been underpinned by the
tremendous promise for potential applications, e.g., for
optoelectronics
\cite{MaierPlasmonicsRev2012,GrigorenkoPlasmonsRev2012}, where the
hydrodynamic approach is particularly suitable for describing the
low-frequency optical response
\cite{DamleSachdev1997,MuellerSachdev2008}. Linearized hydrodynamic
equations provide effective tools for evaluating transport coefficients
in graphene and graphene-based double-layer devices
\cite{PrincipiPlasmons2013,SchuettDrag2013,LevitovDrag2013,GeimSchuettDrag2013,LinHydroBoris2015}.
At the same time, novel experimental techniques
\cite{basov,GrigorenkoPlasmonsRev2012,FeiBasovPlasmonImg2012,ChenKoppensPlasmons2012,abanin,herr,NanoAntennas2014,koppens2015}
bring the studies of nonlinear effects an nonlocal transport phenomena
in graphene is within reach, while improved fabrication methods have
yielded ultra-clean samples \cite{pon}. For example, graphene on
hexagonal boron nitride has been shown to support astonishingly
homogeneous charge densities \cite{Decker2011}.

In this paper we derive a hydrodynamic description of electronic
transport in graphene in the collision-dominated regime, where the
shortest time scale in the problem is provided by electron-electron
interaction. On the contrary, time scales associated with potential
disorder are assumed to be the longest in the system. Consequently,
disorder plays no role in our theory. Our derivation is based on the
quantum kinetic equation (QKE) approach, which has been previously
used to derive the macroscopic linear response theory
\cite{LinHydroBoris2015}.

The transition from the microscopic, kinetic description to the
macroscopic, hydrodynamic equations is simplified by the so-called
``collinear scattering singularity'' of the collision integral
\cite{Kashuba2008,MuellerSachdev2008,SchmalianViscosity2009,Schuett2011,SchuettDrag2013,tom,LinHydroBoris2015}
in the QKE, i.e. the observation that kinematic properties of the
Dirac quasiparticles lead to a divergence in the collision integral
for scattering processes involving quasiparticles moving along the
same direction. Dynamical screening regularizes the divergence
\cite{tom,SchuettDrag2013,LinHydroBoris2015}, such that the resulting
generic relaxation rates in graphene contain a large factor
${\tau_g^{-1}\propto|\ln\alpha_g|\gg1}$, where
${\alpha_g=e^2/\epsilon{v}_g}$ is the effective coupling constant
(here $\epsilon$ is the effective dielectric constant of the substrate
and $v_g$ is the ``speed of light'' in graphene). Depending on the
substrate, the coupling constant may be small
\cite{koz,sheehy,GeimSchuettDrag2013}, ${\alpha_g<1}$. There are,
however, three macroscopic currents \cite{mu12,LinHydroBoris2015} that
are not relaxed at times of order $\tau_g$: (i) the energy current
$\bs{j}_E$; (ii) the electric current $\bs{j}$; and (iii) the
so-called imbalance current \cite{FosterAleiner2009} $\bs{j}_I$.

The energy current $\bs{j}_E$ in graphene is equivalent to the total
momentum of electrons and thus cannot be relaxed by electron-electron
interaction. The electric current in graphene is determined by the
velocity rather than the momentum and therefore is not a conserved
quantity. However, it is conserved in the collinear scattering
processes and hence the corresponding relaxation rate does not contain
the logarithmic enhancement. Finally, the imbalance current
$\bs{j}_I$, is proportional to the sign of the quasiparticle energy
and to the velocity. Similarly to the electric current, it does not
experience logarithmically enhanced relaxation. The imbalance current
is related to the quasiparticle number or imbalance density
\cite{FosterAleiner2009}, ${n_I=n_++n_-}$, where $n_+$ and $n_-$ are
the particle numbers in the upper (conduction) and lower (valence)
bands. Neglecting the Auger processes, quasiparticle recombination due
to e.g. electron-phonon interaction, and three-particle collisions due
to weak coupling, one finds that $n_+$ and $n_-$ are conserved
independently. In this case, which will be considered in the rest of
the paper, not only the total charge density ${n=n_+-n_-}$, but also
the quasiparticle density $n_I$ is conserved.

At times longer than $\tau_g$, physical observables can be described
within the macroscopic -- or hydrodynamic -- approach. The existence
of the three slow-relaxing modes in graphene implies a peculiar
two-step thermalization.

Short-time electron-electron scattering (at time scales up to
$\tau_g$) establishes the so-called ``unidirectional therma-lization''
\cite{SchuettDrag2013}: the collinear scattering singularity implies
that the electron-electron interaction is more effective along the
same direction. Within linear response,\cite{LinHydroBoris2015} one
can express the non-equilibrium distribution function in terms of the
three macroscopic currents $\bs{j}$, $\bs{j}_E$, and $\bs{j}_I$.  The
currents can then be found from the macroscopic equations.  The
currents $\bs{j}$ and $\bs{j}_I$ are not conserved and can be relaxed
by the electron-electron interaction.  Close to charge neutrality, the
corresponding relaxation rates can be estimated as \cite{sheehy,FritzSchmalianCritTransp2008}
${\tau_{ee}^{-1}\sim\alpha^2_g T\ll\tau_g^{-1}}$.  These rates enter
the macroscopic equations as friction-like terms.  The macroscopic
linear response theory has the same form on time scales shorter or
longer than $\tau_{ee}$.

Beyond linear response, the scattering processes characterized by the
time scale $\tau_{ee}$ play an important role in thermalizing
quasiparticles moving in different directions and thus lead to
establishing the local equilibrium. This is the starting point for
derivation of the nonlinear hydrodynamics, which is valid at time
scales much longer than $\tau_{ee}$. In view of conservation of the
particle number, energy, and momentum, as well as independent
conservation of the number of particles in the two bands in graphene,
we may write the local equilibrium distribution function
as\cite{Svintsov2012,Svintsov2013}
\begin{equation}
f^{(0)}_{\lambda,\bs{k}}(\bs{r})
= \left\{1+
\exp\left[\beta(\bs{r})(\ep_{\lambda,\bs{k}}-\mu_\lambda(\bs{r})-\bs{u}(\bs{r})\!\cdot\!\bs{k})\right]
\right\}^{-1}\!,
\label{sec:hydro_micro:eq:local_equilibrium}
\end{equation}
where ${\ep_{\lambda,\bs{k}}=\lambda v_g k}$ denotes the energies of
the electronic states with the momentum $\bs{k}$ in the band
${\lambda=\pm1}$, ${\mu_{\lambda}(\bs{r})}$ the local chemical
potential, the local temperature is encoded in
${\beta(\bs{r})=1/T(\bs{r})}$, and ${\bs{u}(\bs{r})}$ is the
hydrodynamic velocity field which we define below (this field should
not be confused with quasiparticle velocities $\bs{v}$). The
distribution function (\ref{sec:hydro_micro:eq:local_equilibrium})
follows from the standard argument similar to the Boltzmann's
H-theorem \cite{Landau10}: the equilibrium state is characterized by
time-independent entropy. The particular form
(\ref{sec:hydro_micro:eq:local_equilibrium}) takes into account the
symmetry properties of the two-body electron-electron interaction and
is valid for arbitrary single-particle spectrum. The latter means that
Eq.~(\ref{sec:hydro_micro:eq:local_equilibrium}) relies on neither
Galilean nor Lorentz invariance.

Expanding the local equilibrium distribution function
(\ref{sec:hydro_micro:eq:local_equilibrium}) up to the leading order
in deviations from the uniform, equilibrium Fermi distribution, we
recover the distribution function used in the linear response theory
\cite{LinHydroBoris2015}. As we have already mentioned, this
linearized distribution has the same form also on time scales shorter
than $\tau_{ee}$. This is a property of the linear
approximation. Should we attempt to find the subleading nonlinear
terms in the distribution function for ${t<\tau_{ee}}$, the result
would not correspond to the Taylor expansion of
Eq.~(\ref{sec:hydro_micro:eq:local_equilibrium}).

Assuming the local equilibrium
(\ref{sec:hydro_micro:eq:local_equilibrium}) for times
${t\gg\tau_{ee}}$, we derive the nonlinear hydrodynamics in graphene
similarly to the standard Chapman-Enskog procedure
\cite{Landau10,chap,chap2,ens}. The important feature of our theory is
the larger than usual number of hydrodynamic modes (densities of
conserved quantities): total charge, energy, and quasiparticle
imbalance densities and the energy current. The independence of these
modes can be traced to the specific feature of the quasiparticle
spectrum in graphene: the inequivalence of velocity and momentum.

Having derived the hydrodynamic equations, we turn to consider a
representative example of nonlinear physics in graphene, the
relaxation of a hot spot. By this we mean a particular non-equilibrium
state of the system that is characterized by a locally elevated energy
density. Such a state can be prepared with the help of a local probe
or focused laser radiation. Evolving the system according to the
hydrodynamic theory, we find a rather surprising result. Although as
expected \cite{FeiBasovPlasmonImg2012,ChenKoppensPlasmons2012}, the
hot spot emits plasmonic waves that carry energy away, a nonzero
excess energy density remains at the hot spot. Physically, this effect
appears due to compensation between the pressure and the
self-consistent electric (Vlasov) field, which leads to a
quasi-equilibrium. Taking into account the dissipation leads to the
decay of the quasi-equilibrium energy density at the hot spot. This
decay however, is characterized by a longer time scale compared to the
initial emission of the plasmonic waves. At the same time, viscous
effects lead to damping of the plasmonic waves themselves.

The remainder of the paper is organized as follows. In
Sec.~\ref{sec:hydro}, we develop the nonlinear hydrodynamic theory
including dissipative terms starting from the QKE. In
Sec.~\ref{sec:lin_resp} we briefly discuss linear response in
graphene. Finally, Sec.~\ref{sec:nonlin_resp} is devoted to nonlinear
hydrodynamics in graphene. Here we present results on the relaxation
dynamics of a hot spot obtained by a numerical integration of the
hydrodynamic equations. Technical details, e.g., the calculation of
scattering rates for the dissipative terms, are relegated to
appendices.

\section{Hydrodynamic theory in graphene} 
\label{sec:hydro}

In this Section we develop a hydrodynamic theory of transport in
graphene in the collision-dominated regime. We begin with a short
overview of the microscopic mechanisms responsible for establishing
the hydrodynamic regime. The resulting hydrodynamic equations are
summarized in Sec.~\ref{sec:Euler}.

\subsection{From the microscopic theory to hydrodynamics}
\label{sec:hydro_micro}

\subsubsection{Microscopic description}

Microscopically, the electronic system is governed by the Boltzmann
kinetic equation
\begin{equation}
\mathcal{L}f = St_{ee}[f] - \tau_\text{dis}^{-1}(f-\langle f\rangle_\varphi),
\label{sec:hydro_micro:eq:BE}
\end{equation}
with the standard Liouvillian form in the left-hand side,
\begin{equation}
\mathcal{L} = \partial_t + \bs{v}\cdot\bs{\nabla}_{\bs{r}} 
+ \left[e\bs{E}+e(\bs{v}\times\bs{B})\right]\cdot\bs{\nabla}_{\bs{k}}, 
\label{sec:hydro_micro:eq:Liouvillian}
\end{equation}
and the collision integral in the right-hand side. Scattering off
potential disorder is described within the usual $\tau$-approximation with
$\tau_\text{dis}$ being the disorder mean free time. Electron-electron
interaction is described by the collision integral $St_{ee}[f]$.

In graphene, the electronic states can be labeled by the momentum
$\bs{k}$ and the band index ${\lambda=\pm}$. These states are
characterized by the energies
${\epsilon_{\lambda\bs{k}}=\lambda{v}_gk}$ and velocities
${\bs{v}=\lambda{v}_g\bs{k}/k}$, where ${v_g\sim10^6}$m/s. Hereafter
we will work with the units where ${v_g=1}$. Consequently, the
distribution function can be denoted as $f=f_{\lambda\bs{k}}(\bs{r})$.
The angular average in the disorder part of the collision integral
is defined as the average over the direction of $\bs{k}$,
\begin{equation}
\langle f\rangle_\varphi = \int_\varphi f = 
\int\limits_{-\pi}^{+\pi}\!\frac{d\varphi}{2\pi} \: f_{\lambda\bs{k}}.
\label{sec:hydro_micro:eq:angular_average}
\end{equation}

In the interaction-dominated regime, the scattering time
due to electron-electron interaction is much smaller than the
disorder scattering time
\[
\tau_{ee}\gg\tau_\text{dis}.
\]
The same condition was previously used in the derivation of the linear
response theory in graphene. Within linear response, the role of
disorder is to establish the steady state. With the exception of the
charge neutrality point (where in the absence of magnetic field the
steady state can exist without disorder), electron-electron
interaction alone is insufficient for this task. Similarly, in this
paper we keep in mind that infrared divergencies should be cut by
disorder. However, for physical observables, e.g., optical response,
in the frequency window ${\alpha_g^2T\gg\omega\gg\tau_\text{dis}^{-1}}$
the impurity scattering is irrelevant.

We assume that local equilibrium is established at time scales of
the order of ${\tau_{ee}}$, i.e. the longest time scale associated with
two-particle electron-electron interaction. The corresponding length
scale, ${l_\text{hydro}\sim v_g\tau_{ee}}$, defines the size of the
local fluid element \cite{Landau6}. Note, that
${l_\text{hydro}\sim1/(\alpha_g^2T)\gg1/T}$.

Following the standard line of argument \cite{Landau6}, small
deviations from the local equilibrium can be accounted for by
introducing a small correction $\delta f$ to the distribution function
(\ref{sec:hydro_micro:eq:local_equilibrium})
\[
f=f^{(0)}+\delta f.
\]
Kinematic restrictions imposed by the linear spectrum in graphene lead
to the collinear scattering singularity \cite{Arnold2000,Kashuba2008}
in $St_{ee}[f]$. While the singularity is regularized by screening,
most eigenmodes of $St_{ee}[f]$ decay at the shortest time scales
${\tau_g\sim\tau_{ee}/|\ln\alpha_g|}$. As a result, within the leading
logarithmic approximation
\cite{Kashuba2008,FritzSchmalianCritTransp2008} only three modes
contribute to the hydrodynamics and we can parametrize $\delta f$ as
\begin{subequations}
\label{sec:hydro_micro:eq:delta_f}
\begin{equation}
\delta f = T\left(-\frac{\partial f^{(0)}}{\partial\ep_{\lambda\bs{k}}}\right)
\left[\delta f^{(1)} + \delta f^{(2)}\right],
\end{equation}
where
\begin{equation}
\delta f^{(1)} = \frac{v_\alpha}{T} 
\sum_{j=1}^{3}\phi_{j}h_\alpha^{(j)},
\label{sec:hydro_micro:eq:delta_f_1}
\end{equation}
\begin{equation}
\delta f^{(2)} = \frac{v_\alpha v_\beta}{T^2}
\sum_{j=1}^{3}\phi_{j} \: g_{\alpha\beta}^{(j)},
\label{sec:hydro_micro:eq:delta_f_2}
\end{equation}
and the three modes $\phi_j$ are 
\begin{equation}
\phi_1 = 1, \quad
\phi_2 = \lambda, \quad
\phi_3=\ep/T.
\label{sec:hydro_micro:eq:modes}
\end{equation}
\end{subequations}

The non-equilibrium corrections (\ref{sec:hydro_micro:eq:delta_f}) to
the distribution function should leave the conserved quantities
unchanged \cite{Landau10}. As a result, the coefficient ${h^{(3)}=0}$
(which could be understood as a shift of the velocity $\bs{u}$), while
the tensors $g_{\alpha\beta}^{(k)}$ have to be traceless [otherwise
the three terms in Eq.~(\ref{sec:hydro_micro:eq:delta_f_2}) would
shift the particle number density $n$, imbalance density $n_I$, and
energy density $n_E$, respectively].

The coefficients $h_\alpha^{(i)}$ and $g_{\alpha\beta}^{(i)}$ are
determined from the QKE \cite{Arnold2000,FritzSchmalianCritTransp2008},
which becomes a matrix equation in the restricted subspace of modes
$\phi_j$. In what follows, we will use a short-hand notation
\begin{equation}
\label{cmat}
St_{ee}[f]\approx-\mathcal{C}\delta f,
\end{equation}
where the matrix $\mathcal{C}$ corresponds to the linearized collision
integral. The technicalities of inverting the matrix $\mathcal{C}$ are
discussed in Appendix~\ref{sec:app_Coll_Int}, where we also relate the
matrix collision integral to the diagrammatic calculation of
conductivity and viscosity based on the Kubo formula.

Finally, macroscopic equations describing electronic transport in
graphene are obtained by integrating the kinetic equation with the
distribution function (\ref{sec:hydro_micro:eq:delta_f}). In the
hydrodynamic regime, i.e. at time scales much longer than $\tau_{ee}$,
the natural macroscopic variables are the modes that are not relaxed
by electron-electron interaction. All non-conserved quantities should
be expressed in terms of such ``hydrodynamic'' modes. In graphene,
these include the densities $n$, $n_I$, and $n_E$, and the energy
current $\bs{j}_E$. The electric and imbalance currents can then be
found using the equations of state. The emerging hydrodynamics is
valid as long as the macroscopic quantities vary slowly on the scale
$l_\text{hydro}$ set by interactions.

\subsubsection{Macroscopic quantities}

Most two-body electron-electron collisions in graphene leave the
particle number in each band unchanged. This is the consequence of the
linear dispersion relation. The only exception is given by the
so-called Auger processes, where the direction of the momentum of all
initial and final states in each scattering event is the same (i.e.,
all four states belong to the same straight line on the dispersion
cone). In the absence of disorder, the probability of Auger processes
vanishes within the random phase approximation. Even if
impurity-assisted processes are taken into account, the recombination
rate due to Auger processes remains small. Other processes that may
contribute to quasiparticle recombination include electron-phonon
interaction (by means of either two-phonon or impurity assisted
scattering) and three-particle collisions. All these processes
introduce parametrically small relaxation rates
\cite{FosterAleiner2009} (close to charge neutrality, at least of
order ${\alpha_g^4T}$). Here we will neglect recombination and assume
the densities $n_\pm$ to be conserved independently.

The particle and energy densities $n_\pm$ and $n_E$ can be calculated
with the help of the distribution function in a standard way
\begin{subequations}
\begin{equation}
n_+ = \int_{\bs{k}} f_{+,\bs{k}}, 
\label{sec:hydro_macro:eq:n+} 
\end{equation}
\begin{equation}
n_- = \int_{\bs{k}} \left(1-f_{-,\bs{k}}\right), 
\label{sec:hydro_macro:eq:n-} 
\end{equation}
\begin{equation}
n_E = \int_{\lambda,\bs{k}} \ep_{\lambda,k} f_{\lambda,\bs{k}} - n_{E0}.
\label{sec:hydro_macro:eq:nE}
\end{equation}
\end{subequations}
Here we introduced the short-hand notation
\[
\int_{\bs{k}} \dots \equiv N\int\!\frac{d^2 k}{(2\pi)^2} \dots , \quad
\int_{\lambda,\bs{k}} \dots \equiv \sum_{\lambda=\pm} \int_{\bs{k}}  \dots,
\]
where $N=4$ accounts for the spin and valley degeneracy in graphene.
In Eq.~\eqref{sec:hydro_macro:eq:nE} we measure the energy density
$n_E$ with respect to $n_{E0}$,
\begin{equation}
n_{E0}=\int_{\bs{k}}\ep_{-,k}\rightarrow\int_{k<\Delta}\ep_{-,k},
\label{sec:hydro_macro:eq:nE0}
\end{equation}
which is the total energy density at charge neutrality and zero
temperature. The ultra-violet cut-off $\Delta$ must be formally
included [also in Eq.~\eqref{sec:hydro_macro:eq:nE}]. However,
it drops out of the physical results.  

The densities of the conduction and valence bands,
Eq.~\eqref{sec:hydro_macro:eq:n+} and
Eq.~\eqref{sec:hydro_macro:eq:n-}, can be combined into the total
charge and imbalance densities
\begin{subequations}
\begin{equation}
n = n_+ - n_-,
\label{sec:hydro_macro:eq:n}
\end{equation}
\begin{equation}
n_I = n_+ + n_-.
\label{sec:hydro_macro:eq:nI}
\end{equation}
\end{subequations}

The macroscopic currents are defined
\begin{subequations}
\label{js}
\begin{equation}
\bs{j}_+ =  \int_{\bs{k}} \bs{v}_{+,\bs{k}} f_{+,\bs{k}}, 
\label{sec:hydro_macro:eq:j+} 
\end{equation}
\begin{equation}
\bs{j}_- =  \int_{\bs{k}} \bs{v}_{-,\bs{k}} f_{-,\bs{k}},
\label{sec:hydro_macro:eq:j-} 
\end{equation}
\begin{equation}
\bs{j}_E =  \int_{\lambda,\bs{k}} \ep_{\lambda,k}\bs{v}_{\lambda,\bs{k}} f_{\lambda,\bs{k}}.
\label{sec:hydro_macro:eq:jE}
\end{equation}
The electron and hole currents, Eqs.~\eqref{sec:hydro_macro:eq:j+} and
\eqref{sec:hydro_macro:eq:j-} can be combined into the electric and
imbalance (or quasiparticle) currents
\begin{equation}
\bs{j} = \bs{j}_+ - \bs{j}_-,
\label{sec:hydro_macro:eq:j} 
\end{equation}
\begin{equation}
\bs{j}_I = \bs{j}_+ + \bs{j}_-,
\label{sec:hydro_macro:eq:jI}
\end{equation}
\end{subequations}
In graphene the energy current $\bs{j}_E$ is equivalent to the momentum and 
is conserved, while the electric and imbalance currents can be damped by
electron-electron interaction.

\subsection{Generalized Euler equation}
\label{sec:hydro_macro}

In this Section we derive the macroscopic theory of electron transport
in graphene in the absence of dissipation. The resulting hydrodynamic 
equations represent a generalization of the Euler equation of an ideal
liquid to Dirac fermions in graphene.

\subsubsection{Continuity equations in graphene}

The hydrodynamic equations for the densities and currents can be
obtained by averaging the QKE \eqref{sec:hydro_micro:eq:BE} with
respect to the modes~\eqref{sec:hydro_micro:eq:modes}. This yields
the continuity equations for the hydrodynamic densities,
\begin{subequations}
\label{cont_eq}
\begin{equation}
\partial_t n + \bs{\nabla}\cdot\bs{j} = 0,
\label{sec:hydro_macro:eq:cont_n} 
\end{equation}
\begin{equation}
\partial_t n_I + \bs{\nabla}\cdot\bs{j}_I = 0,
\label{sec:hydro_macro:eq:cont_nI} 
\end{equation}
\begin{equation}
\partial_t n_E + \bs{\nabla}\cdot\bs{j}_E = e\bs{E}\cdot\bs{j},
\label{sec:hydro_macro:eq:cont_nE} 
\end{equation}
\end{subequations}
as well as the equation for the energy current
\begin{equation}
\partial_t j_{E,\alpha}+\nabla_\beta\Pi^E_{\beta\alpha}
-e n E_\alpha-en(\bs{u}\times\bs{B})_\alpha
= -j_{E,\alpha}/\tau_\text{dis}.
\label{sec:hydro_macro:eq:eom_jE}
\end{equation}
Using the local distribution function
\eqref{sec:hydro_micro:eq:local_equilibrium}, we can express the
energy current in terms of the hydrodynamic velocity:
\begin{equation}
\bs{j}_E =\frac{3n_E\bs{u}}{2+u^2},
\label{sec:hydro_macro:eq:EOS_jE}
\end{equation}

The equation (\ref{sec:hydro_macro:eq:eom_jE}) includes the momentum
flux or stress tensor
\begin{equation}
\Pi^E_{\alpha\beta}=\int_{\lambda\bs{k}}
\ep_{\lambda,\bs{k}} v_{\alpha} v_\beta f_{\lambda,\bs{k}} \: .
\label{sec:hydro_macro:eq:Pi1}
\end{equation}
In the absence of magnetic field, we use the distribution function
\eqref{sec:hydro_micro:eq:local_equilibrium} and
\eqref{sec:hydro_micro:eq:delta_f} to express $\Pi^E_{\alpha\beta}$ in
terms of $\bs{u}$:
\begin{equation}
\Pi^E_{\alpha\beta}=\frac{n_E}{2+u^2}
\left[\delta_{\alpha\beta}(1-u^2)+3u_\alpha u_\beta\right]
+\delta\Pi^E_{\alpha\beta}.
\label{sec:hydro_macro:eq:EOS_Pi}
\end{equation}
Here the last term $\delta\Pi^E$ describes the dissipative effects
that are considered in the next Section. The first term is the
generalization of the usual stress tensor of an ideal liquid
\cite{Landau6} to the case of Dirac fermions in graphene. The unusual
form of Eq.~(\ref{sec:hydro_macro:eq:EOS_Pi}) reflects the absence of
Galilean as well as Lorentz invariance in the system.

The electric and imbalance currents can be similarly related to the
hydrodynamic velocity
\begin{subequations}
\label{j}
\begin{equation}
\bs{j} = n\bs{u} + \delta\bs{j},
\label{sec:hydro_macro:eq:EOS_j} 
\end{equation}
\begin{equation}
\bs{j}_I = n_I\bs{u} + \delta\bs{j}_I.
\label{sec:hydro_macro:eq:EOS_jI}
\end{equation}
\end{subequations}
Here again we have introduced the dissipative corrections
$\delta\bs{j}$, $\delta\bs{j}_I$. Neglecting these terms along with
$\delta\Pi^E_{\alpha\beta}$, the equations presented in this section
describe the flow of the ideal electronic liquid. Since we are
describing charged particles, the electric field should include the
self-consistent electric (Vlasov) field
\begin{equation}
\bs{E}_{V}(\bs{r})= -\bs{\nabla}_r\!\int\!d^2r' \: V(\bs{r}-\bs{r}') \: \delta n(\bs{r}').
\label{sec:hydro_macro:eq:Vlasov_field}
\end{equation}
Here ${\delta n(\bs{r})=n(\bs{r})-n_0}$ is the local charge
fluctuation, $n_0$ is the background charge density, and
${V(\bs{r})=e^2/r}$ is the 3D Coulomb potential.

\subsubsection{Hydrodynamics of ideal electron liquid}
\label{euler}

In the traditional hydrodynamics \cite{Landau6} the ideal fluid is
described by the Euler equation. The Euler equation is nothing but the
continuity equation for the momentum density, where the stress tensor
is expressed in terms of the velocity field. The latter is typically
done on the basis of Galilean invariance.

Similar equation can be formulated for the electron liquid in
graphene. The momentum density is equivalent to the energy current
which satisfies the continuity
equation~\eqref{sec:hydro_macro:eq:eom_jE}.  Substituting
Eqs.~\eqref{sec:hydro_macro:eq:EOS_jE} and
\eqref{sec:hydro_macro:eq:EOS_Pi} into
Eq.~\eqref{sec:hydro_macro:eq:eom_jE} yields the Euler equation
\begin{equation}
\partial_t \frac{3 n_E u_\alpha}{2+u^2}
+\nabla_\alpha \frac{n_E(1-u^2)}{2+u^2} + \nabla_\beta \frac{3 n_E u_\alpha u_\beta}{2+u^2}
= enE_\alpha
\label{euler_eq}
\end{equation}
This equation is complemented by the continuity equations
(\ref{cont_eq}) and the self-consistency condition
(\ref{sec:hydro_macro:eq:Vlasov_field}). This set of equations
generalizes the hydrodynamics of an ideal liquid to Dirac fermions in
graphene in the absence of dissipation.

\subsection{Dissipative corrections}
\label{sec:hydro_macro_dissipative}

In this Section we extend in the hydrodynamic theory of Dirac fermions
in graphene by taking into account dissipative effects. We use the
explicit form of the non-equilibrium distribution function
(\ref{sec:hydro_micro:eq:delta_f}) to evaluate the dissipative
corrections $\delta\bs{j}$, $\delta\bs{j}_I$, and
$\delta\Pi_{\alpha\beta}$. Comparing our results with the canonical
form of the viscous terms in the stress tensor, we find the expression
for the viscosity coefficients in graphene. We calculate the
dissipative corrections to leading order in the gradient
expansion. The parameter controlling the expansion is similar to the
Knudsen number $\text{Kn}=l_\text{hydro}/l_{\nabla}$, where
$l_{\nabla}$ is the characteristic length scale of hydrodynamic
fluctuations.

\subsubsection{Dissipative corrections to the currents}

Macroscopic equations that describe the electric and imbalance current
densities $\bs{j}$ and $\bs{j}_I$ can be obtained by integrating the
kinetic equation similarly to the derivation of
Eq.~(\ref{sec:hydro_macro:eq:eom_jE}). However, as $\bs{j}$ and
$\bs{j}_I$ are not conserved, the resulting equations contain
non-vanishing contributions of the collision integral. These
contributions can be written in the form
\begin{subequations}
\label{jke}
\begin{equation}
({v},\mathcal{L} f) = -({v},\mathcal{C}\delta f^{(1)}), 
\label{sec:hydro_macro_dissipative:eq:eom_currents_11} 
\end{equation}
\begin{equation}
(\lambda{v},\mathcal{L} f) = -(\lambda{v},\mathcal{C}\delta f^{(1)}),
\label{sec:hydro_macro_dissipative:eq:eom_currents_12}
\end{equation}
where we have used a short-hand notation
\begin{equation}
(g,f) = \int_{\lambda,\bs{k}} g_{\lambda,\bs{k}}f_{\lambda,\bs{k}}.
\label{sec:hydro_macro:eq:scalar_product}
\end{equation}\end{subequations}
Using the distribution function (\ref{sec:hydro_micro:eq:delta_f_1})
we can now construct the explicit relation between the dissipative
corrections to currents and the coefficients $h_{\alpha}^{(j)}$
\begin{subequations}
\begin{equation}
\begin{pmatrix}
\delta j_\alpha \cr
\delta j_{I,\alpha} \cr
\delta j_{E,\alpha}/T 
\end{pmatrix}
= \mathcal{M}
\begin{pmatrix}
h^{(1)}_{\alpha} \cr
h^{(2)}_{\alpha} \cr
h^{(3)}_{\alpha} \cr
\end{pmatrix},
\label{sec:hydro_macro_dissipative:eq:current_to_coeff}
\end{equation}
where the matrix $\mathcal{M}$ is given by
\begin{equation}
\mathcal{M}=\frac{1}{2T}
\begin{pmatrix}
C_1 & C_{\lambda} & C_{\ep/T} \\
C_\lambda & C_1 & C_{|\ep|/T} \\
C_{\ep/T} & C_{|\ep|/T} & C_{\ep^2/T^2}
\end{pmatrix},
\label{sec:hydro_macro_dissipative:eq:matrix_M}
\end{equation}
with the matrix elements
\begin{equation}
C_X = N T \int\limits_{-\infty}^{+\infty}\!
d\ep\nu(\ep)X\left(-\frac{\partial f_0}{\partial\ep}\right).
\label{sec:hydro_macro_dissipative:eq:C_coeff}
\end{equation}
The coefficients $C_{\ep/T}$, $C_{|\ep|/T}$ and $C_{\ep^2/T^2}$ are
proportional to the macroscopic densities
\begin{equation}
C_{\ep} = 2n, \qquad
C_{|\ep|} = 2 n_I, \qquad
C_{\ep^2} = 3n_E/T.
\label{sec:hydro_macro_dissipative:eq:C_coeff_2}
\end{equation}
\end{subequations}
In Eq.~\eqref{sec:hydro_macro_dissipative:eq:C_coeff}, $T$ is the
equilibrium background temperature. 

The relation (\ref{sec:hydro_macro_dissipative:eq:current_to_coeff})
allows us to write the macroscopic equations for the electric and
imbalance currents in the matrix form
\begin{equation}
\partial_t
\begin{pmatrix}
\bs{j} \\
\bs{j}_I
\end{pmatrix}
+ \frac{1}{2}
\begin{pmatrix}
\bs{\nabla} n - e\bs{E} \partial_\mu n \\
\bs{\nabla} n_I - e\bs{E}\partial_\mu n_I
\end{pmatrix}
= -{\mathcal{C}}_J
\begin{pmatrix}
\delta\bs{j} \\
\delta\bs{j}_I
\end{pmatrix}.
\label{sec:hydro_macro_dissipative:eq:eom_currents_2}
\end{equation}
The matrix ${\mathcal{C}}_J$ plays the role of the collision integral
in the reduced three-mode space. Its inverse is given by
\begin{equation}
{\mathcal{C}}^{-1}_J=
\begin{pmatrix}
\tau_{1} & \tau_{2} \\
\tau_{3} & \tau_{4}
\end{pmatrix}.
\label{sec:hydro_macro_dissipative:eq:coll_integral_matrix}
\end{equation}
The transport scattering times $\tau_j$ are obtained from the matrix
elements ${(\phi,\mathcal{C}\phi')}$ of the linearized collision
integral $\mathcal{C}$, where $\phi$ and $\phi'$ are the modes defined
in Eq.~\eqref{sec:hydro_micro:eq:modes}. The off-diagonal times
$\tau_{2,3}$ change their sign for ${n\rightarrow-n}$. In the
non-degenerate regime ${\mu\ll T}$ the times $\tau_j$ are determined
by temperature and electron-electron interaction,
${\tau_j={f_j(\mu/T)}/(\alpha_g^2T)}$, where ${f_j(\mu/T)}$ is a
smooth, dimensionless function. Close to the Dirac point,
\begin{subequations}
\label{taui}
\begin{equation}
\tau_{2}=\tau_{3}=0,
\end{equation} 
while
\begin{equation}
\tau_{1}^{-1} = \frac{\pi}{2T^2\ln 2} ({v}_\alpha,\mathcal{C}{v}_\alpha) 
\approx 2.22 \: \alpha_g^2 T,
\end{equation}
and 
\begin{equation}
\tau_{4}^{-1} = \frac{\pi}{2T^2\ln 2} (\lambda{v}_\alpha,\mathcal{C} \lambda {v}_\alpha) 
\approx 0.05 \: \alpha_g^2 T.
\end{equation}
\end{subequations}
Far away from the Dirac point, ${\mu\gg{T}}$, the system behaves
similarly to the usual Fermi liquid, where the transport mean-free
time due to electron-electron interaction vanishes (physically,
because of the Galilean invariance). Technically, all macroscopic
currents become equivalent and in particular are characterized by the
same transport relaxation rate ${\sim{T}^4/\mu^3}$ which is much
smaller than the usual rate ${\tau_{ee}^{-1}\sim{T}^2/\mu}$
determining both the quasiparticle lifetime and
thermalization. Further details of the calculation are relegated to
Appendix~\ref{sec:app_Gamma_1}.

Solving Eq.~\eqref{sec:hydro_macro_dissipative:eq:eom_currents_2} for
the electric currents to leading order in the gradient expansion
(i.e., in the Knudsen number $\text{Kn}$), we obtain the dissipative
corrections in Eqs.~\eqref{sec:hydro_macro:eq:EOS_j} and
\eqref{sec:hydro_macro:eq:EOS_jI}
\begin{equation}
\begin{pmatrix}
\delta\bs{j} \\
\delta\bs{j}_I
\end{pmatrix}
=
{\mathcal{C}}^{-1}_J\bs{\nu}_J,
\label{sec:hydro_macro_dissipative:eq:diss_corr}
\end{equation}
where the vector $\bs{\nu}_J$ is given by
\begin{equation}
\bs{\nu}_J\!=\!
\begin{pmatrix}
\frac{n}{3n_E}\bs{\nabla} n_E - \frac{1}{2}\bs{\nabla} n - 
\left[\frac{2en^2}{3n_E}-\frac{e}{2}\partial_\mu n\right]\!\bs{E} \\
\frac{n_I}{3n_E}\bs{\nabla} n_E - \frac{1}{2}\bs{\nabla} n_I - 
\left[\frac{2enn_I}{3n_E}-\frac{e}{2}\partial_\mu n_I\right]\!\bs{E}
\end{pmatrix}.
\label{sec:hydro_macro_dissipative:eq:nu_matrix}
\end{equation}
Here we have neglected the frequency dependence formally present in 
Eq.~(\ref{sec:hydro_macro_dissipative:eq:eom_currents_2}) since the
hydrodynamic description is valid at time scales much longer than
the relaxation times due to electron-electron interaction that form
the matrix (\ref{sec:hydro_macro_dissipative:eq:coll_integral_matrix}).

Individual terms in
Eq.~\eqref{sec:hydro_macro_dissipative:eq:nu_matrix} allow for a
simple physical interpretation. The first term in each row describes
the thermoelectric effect; the second term describes diffusion of
electrons and quasiparticles; the last term leads to the finite
conductivity of graphene due to electron interactions
\cite{Kashuba2008}. The latter comprises a Drude-like term, which
becomes more apparent if we identify the mass density $\rho\sim
3n_E/2n$ [see Eq.~\eqref{sec:lin_resp:eq:lin_resp_cond} and the text
  below] and a second term that gives rise to the finite conductivity
at the Dirac point for vanishing charge density $n$.

\subsubsection{Dissipative corrections to the energy stress tensor}

The macroscopic currents (\ref{js}) are defined as the first-order
moments of the distribution function with respect to the three 
modes (\ref{sec:hydro_micro:eq:modes}). The second-order
moments yield the ``generalized stress tensors''
\begin{equation}
	\Pi^{(l)}_{\alpha\beta}
	= \int_{\lambda\bs{k}}\!\phi_l {v}_\alpha {v}_\beta f_{\lambda\bs{k}}.
	\label{sec:hydro_macro_dissipative:eq:def_stress_tensors}
\end{equation}
Here the term with $l=3$ is (up to the factor of $T$) the usual stress
tensor (\ref{sec:hydro_macro:eq:Pi1}). We also define the corresponding
dissipative corrections
\begin{equation}
\delta\Pi^{(l)}=\int_{\lambda\bs{k}}\!\phi_l {v}_\alpha {v}_\beta \delta f_{\lambda\bs{k}}
=
	\begin{cases}
	\delta\Pi , & l=1 \\
	\delta\Pi^I , & l=2 \\
	T^{-1}\delta\Pi^E , & l=3
	\end{cases}
	\label{sec:hydro_macro_dissipative:eq:def_stress_tensor_to_phi}
\end{equation}
where the latter has been already defined in
Eq.~(\ref{sec:hydro_macro:eq:EOS_Pi}).

The dissipative corrections
(\ref{sec:hydro_macro_dissipative:eq:def_stress_tensor_to_phi}) can be
found by integrating the kinetic equation similarly to what was done
for the currents above. This way we find the relation
\begin{equation}
	\begin{pmatrix}
		T \delta \Pi_{\alpha\beta} \\ T \delta
                \Pi^I_{\alpha\beta} \\ \delta \Pi^E_{\alpha\beta} \\
	\end{pmatrix}
	= \frac{1}{2}\mathcal{M}
	\begin{pmatrix}
		g^{(1)}_{\alpha\beta} \\
		g^{(2)}_{\alpha\beta} \\
		g^{(3)}_{\alpha\beta} \\
	\end{pmatrix},
	\label{sec:hydro_macro_dissipative:eq:tensors_to_coeff}
\end{equation}
between $\delta\Pi^{(l)}$ and the coefficients $g_{\alpha\beta}^{(l)}$
from Eq.~\eqref{sec:hydro_micro:eq:delta_f_2}. The matrix
$\mathcal{M}$ is defined in
Eq.~\eqref{sec:hydro_macro_dissipative:eq:matrix_M}. Now we can
express the right-hand side of the integrated kinetic equation in
terms of the $\delta\Pi^{(l)}$. The resulting matrix equation reads
[cf. Eqs.~(\ref{jke}) and
  (\ref{sec:hydro_macro_dissipative:eq:eom_currents_2})]
\begin{equation}
(\phi_l{v}_\alpha{v}_\beta,\mathcal{L}f)
= - (\phi_l{v}_\alpha{v}_\beta,\mathcal{C}\delta f^{(2)}) 
= -{\mathcal{C}}_{\pi,ln}\delta\Pi^{(n)}_{\alpha\beta}.
\label{sec:hydro_macro_dissipative:eq:EOM_tensors_short}
\end{equation}

Inverting the matrix collision integral ${\mathcal{C}}_\pi$, we solve
the above equation and find the dissipative corrections
(\ref{sec:hydro_macro_dissipative:eq:def_stress_tensor_to_phi})
similarly to Eq.~(\ref{sec:hydro_macro_dissipative:eq:diss_corr}):
\begin{subequations}
\label{dpi}
\begin{equation}
\begin{pmatrix}
\delta\Pi_{\alpha\beta} \\
\delta\Pi^I_{\alpha\beta} \\
T^{-1}\delta\Pi^E_{\alpha\beta}
\end{pmatrix}	=
{\mathcal{C}}_\pi^{-1}\nu_{\pi,\alpha\beta},
\label{sec:hydro_macro_dissipative:eq:inverse_coll_int}
\end{equation}
where to leading order in the gradient expansion
\begin{equation}
\nu_{\pi,\alpha\beta}=\frac{1}{4}
\begin{pmatrix}
\delta_{\alpha\beta}\bs{\nabla}\!\cdot\!(n\bs{u}) - \nabla_\alpha n u_\beta - \nabla_\beta n u_\alpha \\
\delta_{\alpha\beta}\bs{\nabla}\!\cdot\!(n_I\bs{u})-\nabla_\alpha n_I u_\beta-\nabla_\beta n_I u_\alpha \\
\frac{3}{2T}[\delta_{\alpha\beta}\bs{\nabla}\!\cdot\!(n_E\bs{u})-\nabla_\alpha n_Eu_\beta-\nabla_\beta n_E u_\alpha]
\end{pmatrix}.
\label{sec:hydro_macro_dissipative:eq:diss_correct_stress_tensor}
\end{equation}
\end{subequations}
The matrix collision integral ${\mathcal{C}}_\pi$ is discussed in
detail in Appendix~\ref{sec:app_Gamma_2}. Hereafter, we restrict our
discussion to the non-degenerate regime, ${\mu\ll{T}}$. Close to the
Dirac point we find
\begin{subequations}
\label{dpidp}
\begin{equation}
\mathcal{C}_\pi=2
\begin{pmatrix}
\mathcal{C}_{\pi,11} & 0 & 0 \\
0 & \mathcal{C}_{\pi,22} & \mathcal{C}_{\pi,23} \\
0 & \mathcal{C}_{\pi,32} & \mathcal{C}_{\pi,33}
\end{pmatrix},
\label{sec:hydro_macro_dissipative:eq:C2_matrix_representation}
\end{equation}
with the matrix elements given by
\begin{equation}
\mathcal{C}_{\pi,ij} = \frac{1}{T}(\phi_i I_{\alpha\beta},\mathcal{C} \phi_k I_{\alpha\beta}) 
(\mathcal{M}^{-1})_{kj}.
\label{sec:hydro_macro_dissipative:eq:C_pi_ij}
\end{equation}
The traceless tensor $I_{\alpha\beta}$ is defined as
\begin{equation}
I_{\alpha\beta} = {v}_\alpha {v}_\beta - \delta_{\alpha\beta}/2.
\label{sec:hydro_macro_dissipative:eq:I}
\end{equation}
\end{subequations}
Close to charge neutrality (see Appendix~\ref{sec:app_Gamma_2} for
details), all matrix elements in
Eq.~(\ref{sec:hydro_macro_dissipative:eq:C_pi_ij}) are of the same
order
\begin{eqnarray}
&&
\frac{1}{T^2}(\ep I_{\alpha\beta},\mathcal{C} \ep I_{\alpha\beta}) 
\sim 
(\lambda I_{\alpha\beta},\mathcal{C} \lambda I_{\alpha\beta})
\\
&&
\nonumber\\
&&
\qquad\qquad\qquad
\sim 
\frac{1}{T}(\lambda I_{\alpha\beta},\mathcal{C} \ep I_{\alpha\beta}) 
\sim \alpha_g^2 T^3.
\nonumber
\end{eqnarray}

The dissipative correction to the stress tensor
(\ref{sec:hydro_macro:eq:EOS_Pi}) is given by the third component of
Eq.~(\ref{sec:hydro_macro_dissipative:eq:inverse_coll_int}).  To
leading order in the fluctuations of the densities, i.e. for $\delta
n_E/n_E\ll1$ as well as $T\delta n_I/n_E\ll1$ and $T\delta n/n_E\ll1$,
the correction $\delta\Pi^E$ takes the canonical form \cite{Landau6}
\begin{equation}
\delta\Pi^E_{\alpha\beta}=
-\eta \left[\nabla_{\alpha} u_\beta + \nabla_{\beta} u_\alpha 
- \delta_{\alpha\beta}\bs{\nabla}\!\cdot\!\bs{u}\right] ,
\label{sec:hydro_macro_dissipative:eq:delta_PI_E}
\end{equation}
with the viscosity coefficient 
\begin{equation}
	\eta = \frac{T}{4} \: \big(\: 0 \: 0 \: 1 \:\big) \: \mathcal{C}_\pi^{-1} \:
		\left(\begin{array}{c}n \\ n_I \\ 3n_E/2T\end{array}\right).
\end{equation}
see Eqs.~\eqref{dpi}. Close to the Dirac point this yields
\begin{equation}
\eta = T(\tau_{\pi,1}n+\tau_{\pi,2}n_I)/4+3\tau_{\pi,3}n_E/8.
\label{sec:hydro_macro_dissipative:eq:viscosity}
\end{equation}
At the Dirac point the first term in
Eq.~\eqref{sec:hydro_macro_dissipative:eq:viscosity} drops out and we
are left with two contributions to the viscosity $\eta$.  The times
$\tau_{\pi,1}$, $\tau_{\pi,2}$ and $\tau_{\pi,3}$ are obtained from
inverting the collision integral \eqref{dpidp} where the charge
density is decoupled from the imbalance and energy densities:
\begin{subequations}
\label{taupi}
\begin{equation}
\tau_{\pi,1} = 0, 
\label{sec:hydro_macro_dissipative:eq:tau_pi_1} 
\end{equation}
\begin{equation}
\tau_{\pi,2} = \frac{1}{2}
\frac{\mathcal{C}_{\pi,32}}
{\mathcal{C}_{\pi,23}\mathcal{C}_{\pi,32}-\mathcal{C}_{\pi,22}\mathcal{C}_{\pi,33}}
\propto \frac{1}{\alpha_g^2T}, 
\label{sec:hydro_macro_dissipative:eq:tau_pi_2} 
\end{equation}
\begin{equation}
\tau_{\pi,3}=\frac{1}{2}
\frac{\mathcal{C}_{\pi,22}}
{\mathcal{C}_{\pi,22}\mathcal{C}_{\pi,33}-\mathcal{C}_{\pi,23}\mathcal{C}_{\pi,32}}
\propto \frac{1}{\alpha_g^2T}.
\label{sec:hydro_macro_dissipative:eq:tau_pi_3}
\end{equation}
\end{subequations}
As a consequence \cite{SchmalianViscosity2009}
\begin{subequations}
\label{eta}
\begin{equation}
\eta(n=0) = B \: T^2/\alpha_g^2,
\end{equation}
where the numerical coefficient is
\begin{equation}
B = \frac{\pi}{12}\:\alpha_g^2T  \tau_{\pi,2}
+\frac{9\zeta(3)}{4\pi}\: \alpha_g^2T\tau_{\pi,3}.
\end{equation}
\end{subequations}
Here we have used the relations $n_E=6\zeta(3)T^3/\pi$ and $n_I=T^2\pi/3$.
Far away form the Dirac point we recover the usual Fermi-liquid 
viscosity \cite{Landau6,prin} ${\eta\propto 1/T^2}$.

Similarly to the classical hydrodynamics \cite{Landau6}, the viscosity
is determined by the homogeneous equilibrium background charge,
imbalance and energy density, or equivalently by the equilibrium
chemical potentials ($\mu_{0,\pm}$) and temperature $T$. The
expression (\ref{sec:hydro_macro_dissipative:eq:delta_PI_E}) implies
vanishing bulk viscosity in graphene. This result is valid within the
leading approximation in the virial expansion that justifies the
kinetic equation (\ref{sec:hydro_micro:eq:BE}) as well as the
distribution function (\ref{sec:hydro_micro:eq:delta_f}).

\subsection{The canonical form of the hydrodynamic equations in graphene} 
\label{sec:Euler}

In this Section we combine the dissipative terms
(\ref{sec:hydro_macro_dissipative:eq:delta_PI_E}) and
(\ref{sec:hydro_macro_dissipative:eq:diss_corr}) with the equations of
the ideal flow in graphene, see Sec.~\ref{euler}. The resulting theory
generalizes the Navier-Stokes hydrodynamics to the Dirac fermions in
graphene.

The complete hydrodynamic description includes the equations of
motion, continuity equations, and equations of state \cite{Landau6}.
Within the local equilibrium approach in graphene, the expression for
the hydrodynamic pressure in terms of the energy density and the
hydrodynamic velocity $\bs{u}$ is highly nonlinear
\begin{subequations}
\label{state_eq}
\begin{equation}
P = \frac{(1-u^2)n_E}{2+u^2}.
\label{sec:Euler:eq:non_linear_pressure}
\end{equation}
For small velocities the pressure assumes the standard value for a
scale invariant gas, ${P_0=n_E/2}$, however, for large velocities
approaching unity ${u\lesssim1}$ it vanishes as ${\sim(1-u^2)}$. The
enthalpy of the system ${W=n_E+P}$ is then given by
\begin{equation}
W = \frac{2w}{2+u^2},\quad
w=n_E+P_0=3n_E/2,
\label{sec:Euler:eq:non_lin_enthalpie}
\end{equation}
\end{subequations}
with the latter being the linear enthalpy of graphene.

The continuity equations (\ref{cont_eq}) are now modified by the
dissipative terms (\ref{sec:hydro_macro_dissipative:eq:nu_matrix}),
\begin{subequations}
\label{cont2}
\begin{equation}
\partial_t n + \bs{\nabla}\cdot(n\bs{u}) = -\bs{\nabla}\cdot\delta\bs{j} ,
\label{sec:Euler:eq:cont_n} 
\end{equation}
\begin{equation}
\partial_t n_I + \bs{\nabla}\cdot(n_I\bs{u}) = -\bs{\nabla}\cdot\delta\bs{j}_I ,
\label{sec:Euler:eq:cont_nI} 
\end{equation}
\begin{equation}
\partial_t n_E + \bs{\nabla}\cdot\left(W\bs{u}\right) = en\bs{E}\cdot\bs{u}.
\label{sec:Euler:eq:cont_nE} 
\end{equation}
\end{subequations}

Finally, adding the dissipative part of the stress tensor
(\ref{sec:hydro_macro_dissipative:eq:delta_PI_E}) to the Euler
equation (\ref{euler_eq}) we obtain a generalization of the
Navier-Stokes equation to Dirac fermions in graphene. Using the
equations of state (\ref{state_eq}), we can bring the resulting
equation to the canonical form (cf.,
Ref~\onlinecite{SchmalianViscosity2009})
\begin{eqnarray}
\label{sec:Euler:eq:Euler_eq_2}
&& 
W\partial_t\bs{u} + W(\bs{u}\cdot\nabla)\bs{u}+\nabla P + \bs{u}\partial_t P 
+ \bs{u}(\delta\bs{j}\cdot\bs{E}) 
\\
&&
\nonumber\\
&&
\qquad\qquad\qquad\qquad\qquad\quad   
= 
en[\bs{E}-\bs{u}(\bs{u}\cdot\bs{E})] + \eta\nabla^2\bs{u}.
\nonumber
\end{eqnarray}
The term ${\bs{u}\partial_t P}$ in the left-hand side of
Eq.~\eqref{sec:Euler:eq:Euler_eq_2} is reflection of nearly
relativistic nature of charge carriers in graphene. In the limit
${u\rightarrow 1}$, the electric field on the
right-hand side of Eq.~\eqref{sec:Euler:eq:Euler_eq_2} does not 
affect the absolute value of the velocity which is limited
by $v_g$.

The complete system of the hydrodynamic equations in graphene includes
Eqs.~(\ref{state_eq}), (\ref{cont2}), (\ref{sec:Euler:eq:Euler_eq_2}),
as well as the equations defining the non-equilibrium corrections to
the electric and imbalance currents
(\ref{sec:hydro_macro_dissipative:eq:diss_corr}).

\section{Linear response} 
\label{sec:lin_resp}

\subsection{Nonlocal optical conductivity}

Evaluation of the linear response transport coefficients within the
hydrodynamic theory is straightforward. Linearizing the Navier-Stokes
equation, we recover the linear response theory derived in
Ref.~\onlinecite{LinHydroBoris2015} with the important addition of
time- and momentum-dependent contributions. Solving these equations,
we find the expression for the momentum-dependent optical conductivity
in graphene up to the subleading order in ${q/\omega}$ [and for
${1/(\omega\tau_{\rm dis}\rightarrow0)}$]
\begin{eqnarray}
\label{sec:lin_resp:eq:lin_resp_cond}
&& 
\sigma(\omega,q) = \sigma_0 + \frac{2i e^2 n^2}{3n_E\omega}
\left[1+\frac{i q^2}{\omega^2}\left(\frac{1}{2}-\frac{2i\eta\omega}{3n_E}\right)\right] 
\\
&&
\nonumber\\
&&
\qquad\qquad\qquad 
+ \frac{i q^2}{\omega}\left[
\frac{\tau_1^2+\tau_2\tau_3}{2}\left(\frac{2e^2n^2}{3n_E}+ e^2 \partial_\mu n\right) 
\right.
\nonumber\\
&&
\nonumber\\
&& 
\qquad\qquad\qquad
\left.
+\frac{\tau_2(\tau_1+\tau_4)}{2}\left(\frac{2e^2n n_I}{3n_E}+ e^2 \partial_\mu n_I\right)
\right].
\nonumber
\end{eqnarray}
Here $\sigma_0$ is the electron-electron contribution to the {\it dc}
conductivity in graphene \cite{Kashuba2008,LinHydroBoris2015}
\begin{subequations}
\label{sigma0}
\begin{eqnarray}
\sigma_0=e^2\!\left[
\tau_1\left(\frac{\partial_\mu n}{2}-\frac{2n^2}{3n_E}\right)
\!+\!\tau_2\left(\frac{\partial_\mu n_I}{2}-\frac{2n n_I}{3n_E}\right)
\right].
\quad\quad
\label{sec:lin_resp:eq:quantum_cond}
\end{eqnarray}
In the above results, $n$, $n_I$ and $n_E$ are the equilibrium
background densities; the scattering times $\tau_i$ follow from
Eq.~\eqref{sec:hydro_macro_dissipative:eq:coll_integral_matrix} (see
also Appendix~\ref{sec:app_Gamma_2}). At the Dirac point, the
electronic compressibility in graphene is
${\partial_\mu{n}=4T\ln2/\pi}$, and hence \cite{hart,mu11}
\begin{equation}
\label{sdp}
{\sigma_0=Ae^2/\alpha_g^2}, 
\end{equation}
\end{subequations}
where we find ${A=0.19}$ (previously, the value ${A=0.12}$ was
reported in Ref.~\onlinecite{Kashuba2008}).

At ${q=0}$, the conductivity
\eqref{sec:lin_resp:eq:lin_resp_cond} can be interpreted in
terms of the usual Drude formula, where the role of the effective mass
density is played by the ratio ${3n_E/(2n)}$.

The result \eqref{sec:lin_resp:eq:lin_resp_cond} suggests a
possibility to measure the viscosity coefficient in graphene in
nonlocal transport measurements \cite{abanin,herr}. However, precisely
at the Dirac point (${n=0}$) the optical conductivity is independent
of viscosity. Physically, viscosity is associated with the momentum
density, i.e. the energy current. At the Dirac point, the energy and
electric currents decouple \cite{LinHydroBoris2015} and hence the
conductivity is unaffected by viscous effects.

Finally, let us remark on the apparent contradiction between
Eq.~\eqref{sec:lin_resp:eq:lin_resp_cond} and the corresponding result
of Ref.~\onlinecite{MuellerSachdev2008}, where it was found that the
expansion of the optical conductivity in ${q/\omega}$ contains linear
terms missing in Eq.~\eqref{sec:lin_resp:eq:lin_resp_cond}. The reason
for this disagreement is that we have calculated the response to the
total electromagnetic field, while the result of
Ref.~\onlinecite{MuellerSachdev2008} represents the response to the
external field. In the latter case one has to take into account
screening which leads to the linear in $q$ terms in nonlocal
conductivity.

\subsection{Hydrodynamic energy waves and plasmons}

In a formally infinite system, the hydrodynamic theory
(\ref{state_eq}) - (\ref{sec:Euler:eq:Euler_eq_2}) admits solutions in
the form of collective energy waves with the dispersion relation
(which we obtain as an expansion in ${q/\omega<1}$)
\begin{eqnarray}
\label{sec:lin_resp:eq:disp_rel}
&&
\omega(q)= -\frac{i}{2\tau_{\rm dis}} + i\pi q \frac{\alpha_g\sigma_0}{e^2}
- iq^2\left(\frac{\eta}{n_E}+ \frac{\tau_1+\tau_4}{4}\right)
\nonumber\\
&&
\nonumber\\
&&
\qquad
+\Bigg[\frac{q^2}{2}
\left(1+\frac{4\alpha_g n^2}{3n_E q}\right) 
\\
&&
\nonumber\\
&&
\qquad
-q^4
\left(\frac{\eta}{3n_E}-\frac{\sigma_0}{2e^2}\frac{\alpha_g}{q} 
+ \frac{\tau_1+\tau_4}{4} + \frac{1}{2\tau_{\rm dis} q^2}\right)^2
\Bigg]^{\frac{1}{2}}\!\!.
\nonumber	
\end{eqnarray}
These solutions can be interpreted as the hydrodynamic zero modes
corresponding to poles in the response functions, see
Appendix~\ref{sec:app_lin_response}. Here we have also taken into
account weak disorder, which is absent in
Eq.~(\ref{sec:Euler:eq:Euler_eq_2}).

For pure systems in the absence of dissipation the dispersion relation
(\ref{sec:lin_resp:eq:disp_rel}) greatly simplifies. At charge
neutrality (${n=0}$), the leading term is linear in $q$,
\begin{equation}
\label{cosmic}
\omega(n=0, \tau_{\rm dis}\rightarrow\infty, \eta\rightarrow 0) \approx v_g q /\sqrt{2}.
\end{equation}
This acoustic energy wave \cite{levitov} is analogous to the long-wavelength
oscillations in interacting systems of relativistic particles \cite{Landau6},
sometimes called ``cosmic sound''. Such oscillations play an important role 
in astrophysics \cite{as1,as2}.

\begin{figure}[t]
\centerline{\includegraphics[width=0.9\linewidth]{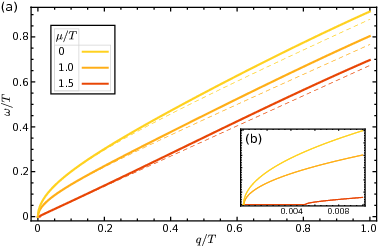}}
\caption{(Color online) Energy wave dispersion
  \eqref{sec:lin_resp:eq:disp_rel}, for different chemical
  potentials. In the main panel (a), we compare the energy waves in an
  ideal fluid (dashed lines) to the dissipative (viscous) flow (solid
  lines). The inset (b) shows the effect of disorder scattering at
  small momenta. The curves are calculated for
  ${1/T\tau_\text{dis}=0.001}$.}
\label{sec:lin_resp:fig:plasmon_disp}
\end{figure}

Away from charge neutrality, the collective modes of a pure system exhibit
the square root spectrum typical for 2D plasmons:
\begin{equation}
\label{plas}
\omega(\tau_{\rm dis}\rightarrow\infty, \eta\rightarrow 0) \approx
n\sqrt{\frac{2\alpha_gq}{3n_E}}.
\end{equation}
Let us stress, that this mode is {\it not} the usual RPA plasmon. The
crucial point is that the hydrodynamic description developed in this
paper is valid at length scales much longer than the scale
${l_{\rm{hydro}}}$, associated with electron-electron interaction,
i.e. for very small momenta ${q\ll{l}^{-1}_{\rm hydro}}$. In contrast,
the usual RPA plasmons \cite{Schuett2011,PrincipiPlasmons2013} are
discussed for momenta that are large compared to the characteristic
scales of both disorder and interaction.

In a regular 2D electron systems, electric current is relaxed by
disorder and as a result, the plasmon waves are damped at the lowest
momenta. The plasmon dispersion is given by \cite{zna}
\[
\omega \left(\omega + \frac{i}{\tau_{\rm dis}}\right) 
= \frac{1}{2} \varkappa q v^2_F,
\]
such that for momenta smaller than the inverse Thomas-Fermi screening
radius 
\begin{equation}
\label{pl2dd}
\omega(q\ll\varkappa) =
-\frac{i}{2\tau_{\rm dis}} + \sqrt{\frac{1}{2} \varkappa q v^2_F -
  \frac{1}{4\tau^2_{\rm dis}}}.
\end{equation}
As a result, for momenta much smaller than the inverse mean-free path
the plasmon dispersion is purely imaginary, as expected for diffusive
systems. For energy waves in graphene disorder scattering plays a
similar role, see Eq.~(\ref{sec:lin_resp:eq:disp_rel}).

Moreover, in graphene the electric current is relaxed also by
electron-electron interactions
\cite{DasSarmaScreening2007,Kashuba2008,FritzSchmalianCritTransp2008,Schuett2011,SchuettDrag2013,plarev,LinHydroBoris2015}.
As a result, the plasmon modes are damped \cite{plarev} similarly to
Eq.~(\ref{pl2dd}) even in the absence of disorder:
\[
\omega = -\frac{i}{2\tau_{ee}} + \sqrt{\omega_p^2 -
  \frac{1}{4\tau^2_{ee}}},
\]
where ${\omega_p^2=\varkappa{q}/2}$ for ${q\ll\varkappa}$ with the
inverse Thomas-Fermi screening radius being
${\varkappa=2\pi\alpha_g(\partial_\mu{n})}$. Such plasmons exist even
at charge neutrality \cite{Schuett2011} (for ${T>0}$). Thus for small
momenta, the plasmons are overdamped in contrast to the energy waves
(\ref{sec:lin_resp:eq:disp_rel}). However, away from charge
neutrality, the energy waves hybridize with the charge sector due to
Vlasov self-consistency leading to dynamic oscillations of the charge
density with the dispersion (\ref{plas}), that is similar to
$\omega_p$, but with a smaller prefactor. These oscillations should be
experimentally observable in the same way as usual plasmons
\cite{FeiBasovPlasmonImg2012,ChenKoppensPlasmons2012}, provided that
the samples (as well as the time scale of the measurements) are in the
hydrodynamic regime.

Far away from the Dirac point (${\mu\gg T}$), the distinction between
the charge and energy sectors of the theory disappears, such that the
energy waves coincide with the usual plasmon \cite{levitov}: for
${\mu\gg T}$, the dispersion (\ref{plas}) reproduces $\omega_p$.
Technically, the transport relaxation time due to electron-electron
interaction that determines the above plasmon damping becomes much
longer than the usual electron-electron scattering time that is
responsible for thermalization in the system, see discussion following
Eqs.~(\ref{taui}).

Viscous forces influence the collective modes
(\ref{sec:lin_resp:eq:disp_rel}) in the higher order in ${q/\omega}$,
cf. Eq.~(\ref{sec:lin_resp:eq:lin_resp_cond}). Unlike the case of the
optical conductivity, here viscosity enters in a linear combination
with the scattering times $\tau_1$ and $\tau_4$. Consequently,
measuring the energy wave dispersion might not be the best way to find
the viscosity in graphene. However, combining such measurements with
the measurement of nonlocal conductivity, one can find experimental
values for not only $\eta$, but also the scattering times $\tau_i$.

The above results are illustrated in
Fig.\ref{sec:lin_resp:fig:plasmon_disp}, where we plot the dispersion
(\ref{sec:lin_resp:eq:disp_rel}) for different chemical potential. The
inset illustrates the role of disorder, cf. Eq.~(\ref{pl2dd}).

\begin{figure*}[t]
\centerline{\includegraphics[width=0.9\linewidth]{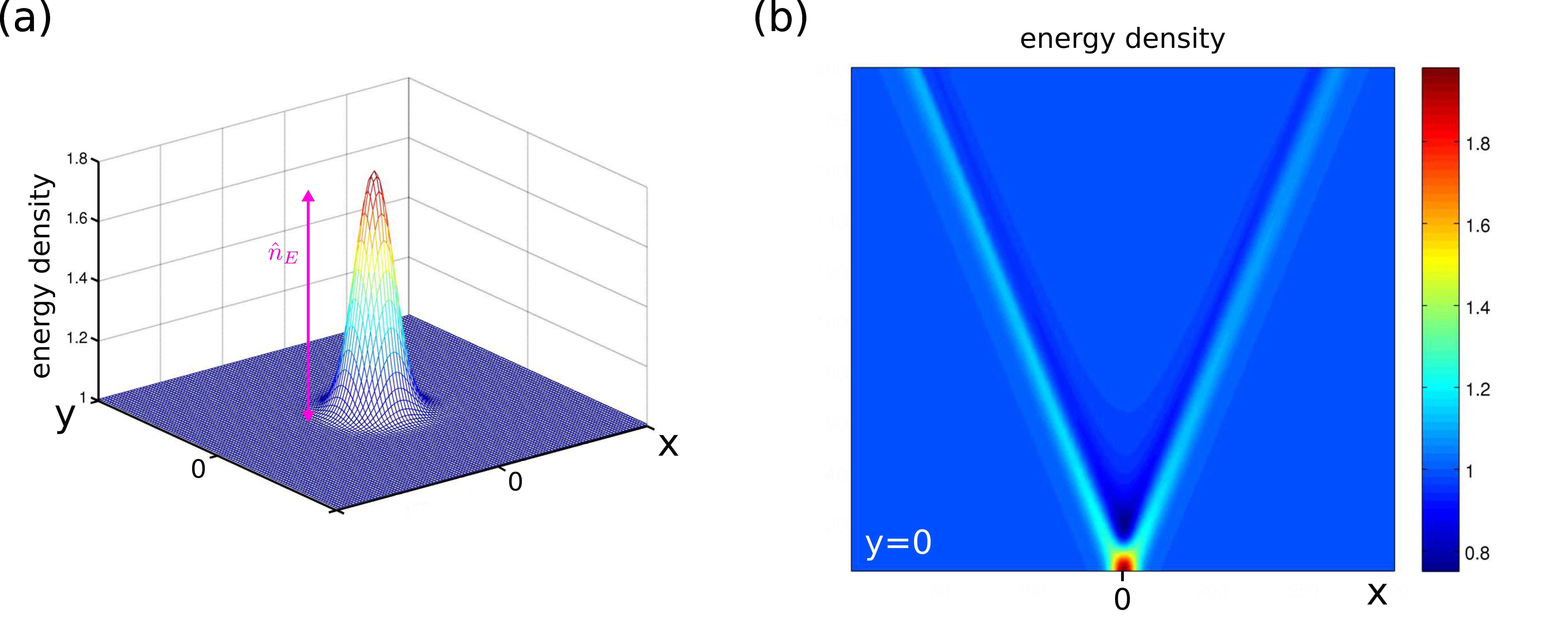}}
\caption{(Color online) Hot spot relaxation of a neutral ideal fluid
  obtained from the Euler hydrodynamics (\ref{cont_eq}) -
  (\ref{euler_eq}) without the Vlasov self-consistent electric field,
  ${\bs{E}=0}$. The left panel shows the initial energy bump with the
  height ${n_{E}=1.8n_E^{(0)}}$. The right panel shows the evolution of
  the energy density (in units of the equilibrium background,
  ${n_E/n_E^{(0)}}$) as a function of the $x$-coordinate and time
  (arbitrary units) along the line ${y=0}$.}
\label{sec:nonlin_resp:fig:hotspot_dissipationless}
\end{figure*}

\section{Nonlinear effects: relaxation of a hot spot} 
\label{sec:nonlin_resp}

In this Section we report results of a numerical integration of the
nonlinear hydrodynamic equations (\ref{state_eq}) -
(\ref{sec:Euler:eq:Euler_eq_2}) describing relaxation of a hot spot.

Let us prepare the system in a homogeneous, equilibrium state
characterized by the charge density $n^{(0)}$ (i.e., away from charge
neutrality), energy density $n_E^{(0)}$ and imbalance density
$n_I^{(0)}$. On top of this equilibrium background, we create a hot
spot: a locally elevated energy density. For simplicity, we choose a
Gaussian profile with the peak height ${n_{E}=1.8n_E^{(0)}}$, see
Fig.~\ref{sec:nonlin_resp:fig:hotspot_dissipationless}(a). The
resulting non-equilibrium state will serve an initial condition for the
subsequent time evolution that follows Eqs.~(\ref{state_eq}) -
(\ref{sec:Euler:eq:Euler_eq_2}).

The computer simulations are performed in a semi-implicit scheme
\cite{AscherIMEX1995}. The diffusive and viscous corrections are
discretized implicitly. This scheme is suitable for a wider class of
problems that are characterized by competing convective and diffusive
terms. Moreover, the simulations are performed on a staggered grid to
avoid unphysical density oscillations \cite{GriebelFluidBook}.

\subsection{Ideal flow}

We begin with the evolution of the hot spot in an ideal system
described by the Euler hydrodynamics (\ref{cont_eq}) -
(\ref{euler_eq}). Here we assume that the system is not subjected to
any external fields.

\subsubsection{Pure energy flow}

Within the hydrodynamic approach, the energy flow is coupled to the
charge flow by means of the self-consistent electric field
\eqref{sec:hydro_macro:eq:Vlasov_field}. Turning off the Vlasov
terms (i.e., setting ${\bs{E}=0}$), we arrive at an essentially
neutral system where the energy flow is decoupled from the rest
of the degrees of freedom.

In such a system, creating an excess energy density leads to
excitation of ballistic (due to absence of dissipation) energy waves
with the linear dispersion (\ref{cosmic}). This flow is illustrated
in Fig.~\ref{sec:nonlin_resp:fig:hotspot_dissipationless}(b), where we
plot the resulting energy density profile along the line ${y=0}$ as a
function of the $x$-coordinate and time. In
Fig.~\ref{sec:nonlin_resp:fig:hotspot_dissipationless} we use
arbitrary units, since the time and length scales associated with the
ballistic propagation in an ideal system are determined by the initial
conditions.

The decay of the hot spot into the energy waves does not lead to an
immediate relaxation of the initial energy density profile, see
Fig.~\ref{sec:nonlin_resp:fig:hotspot_dissipationless}(b). In contrast
to the three-dimensional flow, the Green's function of the 2D wave
equation exhibits a long-time tail, ${\sim t^{-1}}$. As a consequence
the relaxation of the hot spot in the dissipationless limit without
Vlasov field shows power law decay. This slow relaxation of the energy
density around the origin (afterglow) can be seen in
Fig.~\ref{sec:nonlin_resp:fig:hotspot_dissipationless}(b).

\subsubsection{Charge fluctuations}

\begin{figure}[t]
\centerline{\includegraphics[width=0.9\linewidth]{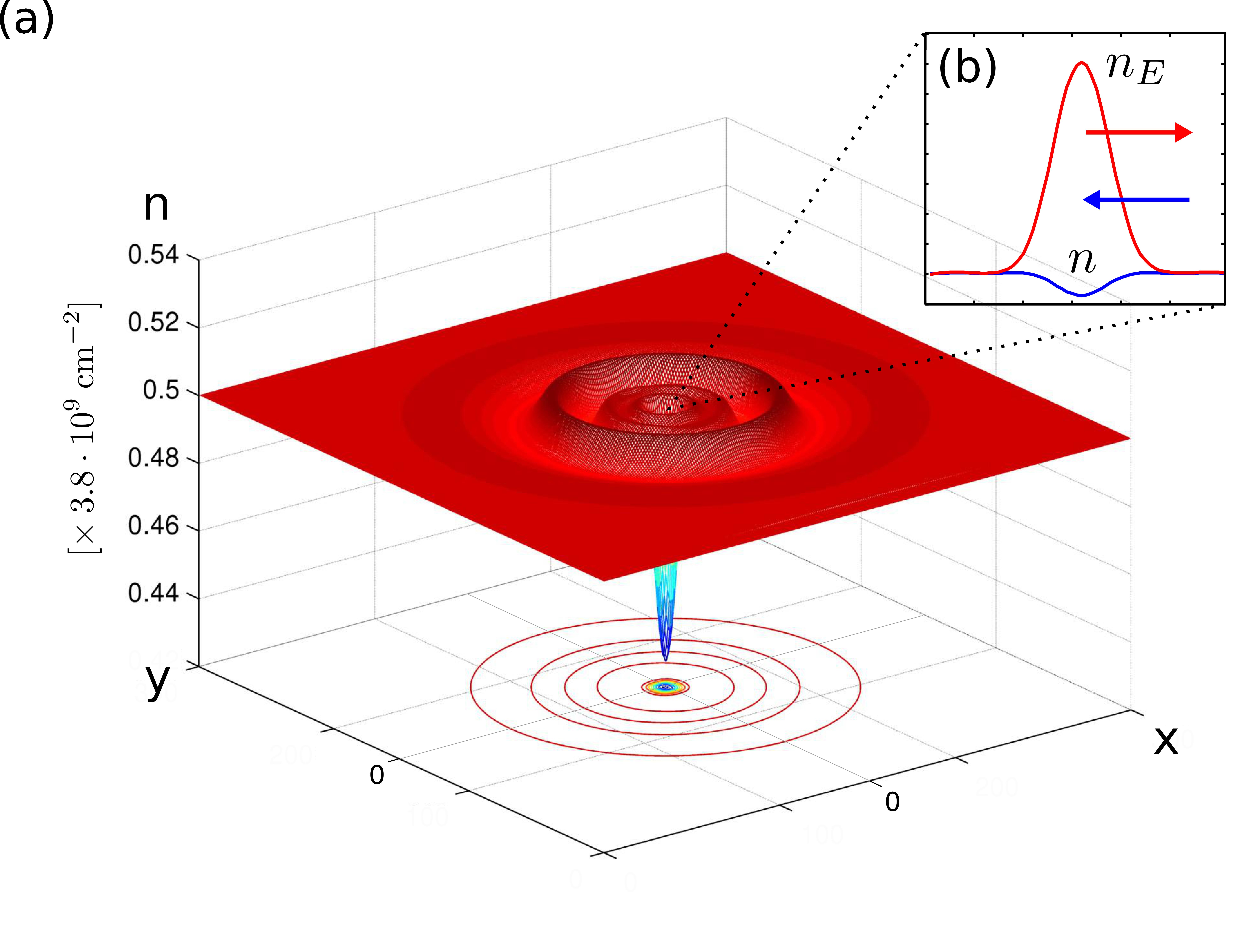}}
\caption{(Color online) A snapshot of the charge density $n$. The
  equilibrium value of the charge density is
  ${n^{(0)}=1.9\times10^{9}\text{cm}^{-2}}$. The initial height of the
  energy bump is ${n_{E}=1.8n_E^{(0)}}$. The inset (b) illustrates the
  soliton-like composite profile that is established at the
  origin. The blue curve shows the dip in the charge density and the
  red curve shows the excess energy density. The arrows show the
  balanced hydrodynamic forces: the pressure (red arrow) and the
  self-consistent electric field (blue arrow).}
\label{sec:nonlin_resp:fig:hotspot_simul}
\end{figure}

In a charged system, i.e., in the presence of the self-consistent
electric field, the cosmic sound wave shown in
Fig.~\ref{sec:nonlin_resp:fig:hotspot_dissipationless} is accompanied
by fluctuations of the charge density, see
Fig.~\ref{sec:nonlin_resp:fig:hotspot_simul}.

The excess energy density generates the pressure force described by
${\nabla_\beta\Pi^E_{\beta\alpha}}$. This creates the initial energy
flow that corresponds to the nonzero hydrodynamic velocity $\bs{u}$,
see Eq.~(\ref{sec:hydro_macro:eq:EOS_jE}) and hence translates into an
electric current (\ref{sec:hydro_macro:eq:EOS_j}), which is coupled to
the charge density by means of the continuity equation
(\ref{sec:hydro_macro:eq:cont_n}). This way, the initial evolution of
the excess energy density leads to a depletion of the charge density
at the origin.

Now, the non-equilibrium charge density profile results in the
self-consistent electric field [due to Vlasov terms
  (\ref{sec:hydro_macro:eq:Vlasov_field})]. Remarkably, in the absence
of dissipation the electric field partially compensates the pressure
force leading to the appearance of a stable soliton-like composite
density profile at the origin: after the initial outflow of energy
carried away by the cosmic sound waves, some excess energy density
remains at the point of the initial perturbation accompanied by the
dynamically generated dip in the charge density, see
Fig.~\ref{sec:nonlin_resp:fig:hotspot_simul}.

The establishing of the depletion in the charge density is accompanied
by the charge flow shown in
Fig.~\ref{sec:nonlin_resp:fig:hotspot_simul_2}. Although that figure
shows the flow in the presence of dissipation, at the short time
scales used in the figure the dissipative effects are still weak and
the resulting flow can be considered dissipationless.

\subsection{Dissipative relaxation dynamics}
\noindent

Consider the hot spot relaxation in a fully interacting system,
i.e. in the presence of dissipation. We start with the same initial
condition as before, but now the system evolves under the
Navier-Stokes hydrodynamics (\ref{state_eq}) -
(\ref{sec:Euler:eq:Euler_eq_2}).

\begin{figure}[t]
\centerline{\includegraphics[width=0.9\linewidth]{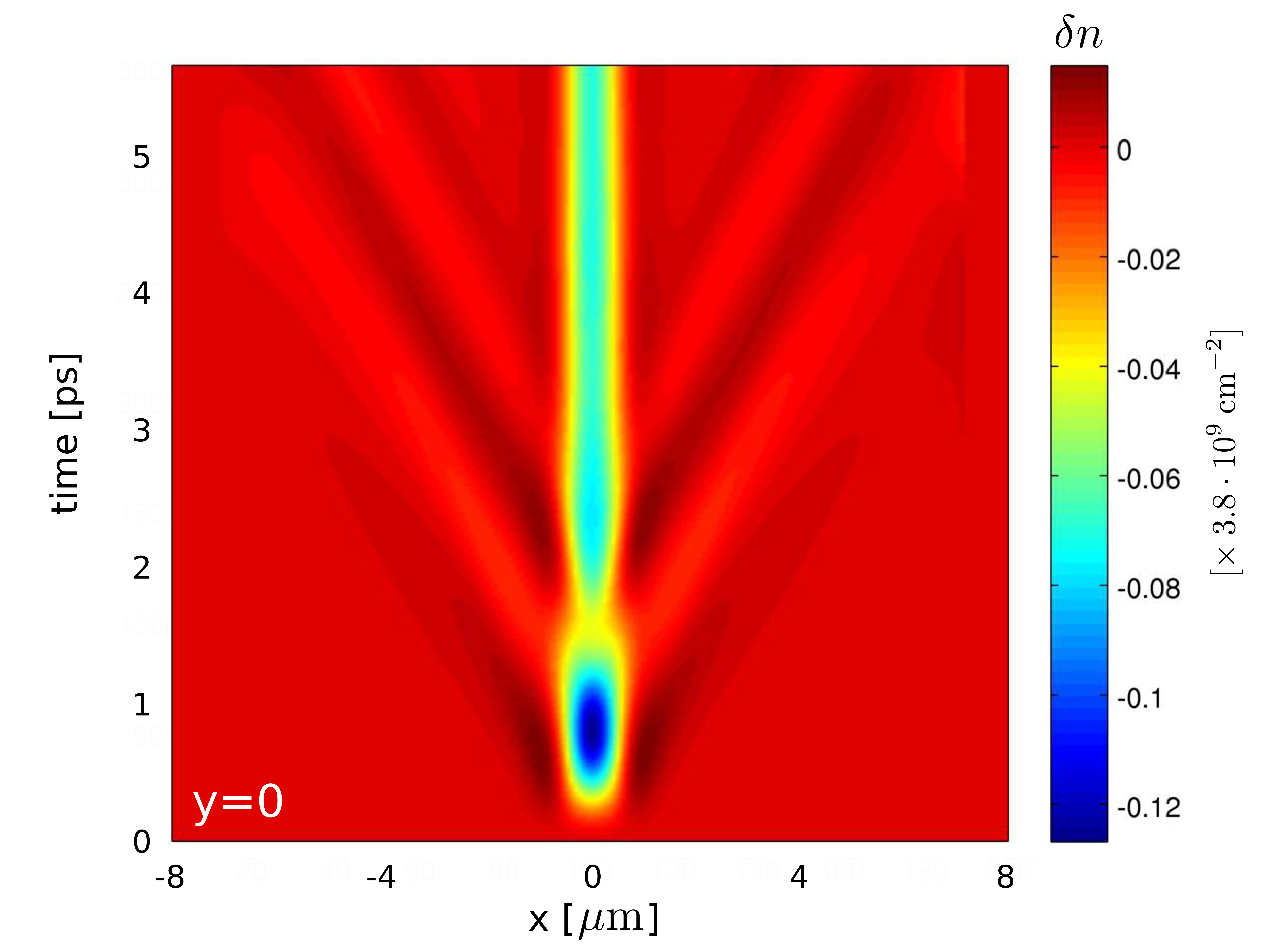}}
\caption{(Color online) The charge density as a function of $x$ along
  the line $y=0$ for short enough time scales such that the system is
  effectively in the dissipationless limit.}
\label{sec:nonlin_resp:fig:hotspot_simul_2}
\end{figure}

The hot spot evolution now proceeds in two stages. The first stage is
similar to the ideal flow, where the quasi-stable charge-energy
density profile is established at the origin. During this stage, some
energy and charge are being carried away from the hot spot by the
emitted energy waves. The metastable patterns, such as the
charge-energy complex in
Fig.~\ref{sec:nonlin_resp:fig:hotspot_dissipationless}(b) and the
traveling waves are formed due to the nonlinear interplay between the
charge and energy sectors. These patterns were stable in the absence
of dissipation, but now acquire a finite lifetime.

Dissipative effects are characterized by a distinctly longer time
scale compared to the initial evolution of the hot spot. These effects
are manifested during the second stage of the hot spot evolution. Here
the electron-electron interaction leads to damping of the emitted
waves, with the damping rate given by the imaginary part of the
spectrum (\ref{sec:lin_resp:eq:disp_rel}). In the clean limit, the
dominant contribution to the damping rate is linear in $q$ (similar to
the 2D Maxwell relaxation, but with $\sigma_0$ determined by
electron-electron interaction). Furthermore, the soliton-like
charge-energy complex is no longer stable and decays. However, the
depletion of the charge density at the origin remains visible for at
least several picoseconds after the initial perturbation, see
Fig.~\ref{sec:nonlin_resp:fig:hotspot_simul_2} and hence should be
detectable by modern experimental techniques
\cite{FeiBasovPlasmonImg2012,ChenKoppensPlasmons2012}.

\section{Conclusions} 
\label{sec:conclusion}

In this paper we have presented a hydrodynamic description of the
electronic transport in graphene. Our formalism allows for a
consistent treatment of nonlinear hydrodynamic effects as well as
dissipative phenomena due to electron-electron interaction. Our theory
describes the following hydrodynamic modes: the energy, particle and
imbalance densities and the energy current. The electric and imbalance
currents are relaxed by electron-electron scattering and have to be
obtained from the equations of state. The resulting macroscopic
description includes a generalization of the Navier-Stokes equation in
graphene (\ref{sec:Euler:eq:Euler_eq_2}), the nonlinear relations
(\ref{state_eq}) between the hydrodynamic pressure and enthalpy and
the hydrodynamic velocity $\bs{u}$ that is related to the energy
current. These relations play the role of the equations of
state. Finally, the three macroscopic densities obey the set of
continuity equations (\ref{cont2}).

Having derived the hydrodynamic theory from the Boltzmann kinetic
equation, we are able to calculate explicitly the set of scattering
times that determine the coefficients in the hydrodynamic equations,
in particular the viscosity (\ref{eta}) and the {\it dc}-conductivity
at charge neutrality (\ref{sdp}). The latter is the manifestation of
the non-Galilean-invariant nature of the electronic system in
graphene, where the electric current can be relaxed by
electron-electron interaction.

In laboratory experiments, viscous effects can be detected, for
instance, by measuring nonlocal conductivity in graphene
\cite{abanin,herr}. Within linear response, viscosity affects the
conductivity away from charge neutrality and at nonzero
momenta. Another experimentally detectable viscous effect is the
plasmon lifetime in graphene. Although the viscosity coefficient
enters the plasmon damping in a linear combination with other
interaction-dependent parameters, see
Eq.~(\ref{sec:lin_resp:eq:disp_rel}), measuring both the plasmon
lifetime and nonlocal conductivity may give experimental access to
several relaxation times determined by electron-electron interaction.

Beyond linear response, we have considered the simplest example of
nonlinear phenomena in graphene - the relaxation dynamics of a hot
spot, see Fig.~\ref{sec:nonlin_resp:fig:hotspot_dissipationless}. This
analysis takes into account the convective nonlinearities and the
residual Coulomb interaction. In the macroscopic equations, the latter
manifests the self-consistent electric field due to charge
fluctuations and the dissipative corrections. We have found that the
hot spot relaxation proceeds in two stages. The first stage, lasting
no longer than few picoseconds, is characterized by the metastable
charge-energy profile at the origin and the traveling energy waves
that carry excess energy and charge away from the hot spot, see
Figs.~\ref{sec:nonlin_resp:fig:hotspot_simul},
\ref{sec:nonlin_resp:fig:hotspot_simul_2}. The emitted waves exhibit
characteristic modulation due to the self-consistent Vlasov electric
field. During the second stage, dissipative effects start playing a
definitive role in the process leading to he diffusive charge
propagation, damped energy waves, and the decay of the soliton-like
charge-energy profile at the origin.  The dissipative effects are much
slower than the initial evolution of the hot spot. In particular, the
metastable charge-energy profile remains visible at times of order
$10$ps, which should be detectable in laboratory, see
Fig.~\ref{sec:nonlin_resp:fig:hotspot_simul_2}.

The traveling energy waves are accompanied by fluctuations of the
charge density due to nonlinear coupling between the energy and charge
sectors in the theory away from charge neutrality. Precisely at the
Dirac point, the energy waves have linear dispersion (\ref{cosmic}),
similar to the cosmic sound \cite{levitov}. For finite background
charge densities the dispersion of the energy waves (\ref{plas})
becomes similar to the usual 2D plasmons \cite{PrincipiPlasmons2013},
with its intrinsic life-time determined by electron-electron
interaction. However, as the hydrodynamic theory is valid only for
time and length scales that are much larger than the typical scales
associated with the electron-electron scattering, the true plasmon
modes remain overdamped \cite{plarev}. However, far away from charge
neutrality (${\mu\gg T}$) we recover the usual plasmon in graphene.

The hydrodynamic theory presented in this paper is valid as long as
quasiparticle recombination processes remain slow (technically,
infinitely slow). At time scales exceeding the recombination times the
imbalance density is no longer conserved and the structure of the
hydrodynamic equations changes. However, the Navier-Stokes equation
(\ref{sec:Euler:eq:Euler_eq_2}) is independent of the imbalance
density and remains valid even at the longest time scales.

The problem of the hot spot relaxation and traveling energy waves
considered in this paper is closely related to recent experimental
imaging of plasmons in graphene
\cite{FeiBasovPlasmonImg2012,ChenKoppensPlasmons2012,koppens2015}.
While the existing experiments are focusing on the high-frequency
optical phenomena, we hope that our investigation of the energy waves
in graphene will motivate future measurements in the low-frequency,
hydrodynamic regime. At the same time, nonlocal transport measurements
\cite{abanin,herr} may uncover exciting manifestations of the
nonlinear, viscous flow in graphene including vortices and laminar
wake.

Our hydrodynamic theory can be further applied to more realistic,
experimentally relevant geometries in order to study possible
realizations of the plethora of hydrodynamic phenomena in graphene.
After a straightforward generalization, the theory allows us to
consider the thermoelectric effects as well as the effects of the
external magnetic field. This work will be reported elsewhere.

\begin{acknowledgments}
\noindent
We would like to thank I.A. Dmitriev, M.I. Katsnelson, A. Levchenko,
L.S. Levitov, J. Schmalian, and L.A. Ponomarenko for very fruitful
discussions. Furthermore we want to thank C. Seiler for his invaluable
help with computer simulations. This work was supported by the EU
Network Grant InterNoM, DFG SPP 1459 and by the Alexander-von Humboldt
Stiftung.
\end{acknowledgments}

\appendix

\onecolumngrid

\section{The ee-collision integral} 
\label{sec:app_Coll_Int}

The electron-electron collision integral in the QKE
\eqref{sec:hydro_micro:eq:BE} is given by
\begin{subequations}
\label{colint}
\begin{eqnarray}
&& 
St_{ee}[f]=\sum_{\nu,\lambda',\nu'}\int_{\bs{k},\bs{p}',\bs{k}'}
|M|^2 (2\pi)^3
\delta(\ep_{\lambda p}+\ep_{\nu k}-\ep_{\lambda' p'}-\ep_{\nu' k'})
\delta(\bs{p}+\bs{k}-\bs{p}'-\bs{k}') 
\\
&&
\nonumber\\
&& 
\qquad\qquad \qquad\qquad \qquad\qquad 
\times 
\left\{
f_{\lambda',\bs{p}'} f_{\nu',\bs{k}'} \left[1-f_{\nu,\bs{p}}\right] 
\left[1-f_{\lambda,\bs{k}}\right]
-
f_{\lambda,\bs{p}} f_{\nu,\bs{k}} \left[1-f_{\nu',\bs{p}'}\right] 
\left[1-f_{\nu',\bs{k}'}\right]
\right\}.
\nonumber
\label{sec:app_Coll_Int:eq:coll_int}
\end{eqnarray}
Here the matrix element of Coulomb scattering is given by
\begin{equation}
|M|^2=N |V(\omega,q)|^2 \Theta_{\lambda \bs{p}, \lambda' \bs{p}'}
\Theta_{\nu \bs{k}, \nu' \bs{k}'},
\label{sec:app_Coll_Int:eq:Coulomb_ME}
\end{equation}
with the graphene specific Dirac factors
\begin{equation}
\Theta_{\lambda,\bs{p};\lambda',\bs{p}'}
= \frac{1}{2}\left(1+\lambda\lambda'\frac{\bs{p}\cdot\bs{p}'}{pp'}\right)
= \frac{1}{2}\left(1+\hat{v}_{\lambda,\bs{p}}\cdot\hat{v}_{\lambda',\bs{p}'}\right),
\label{sec:app_Coll_Int:eq:Dirac_factor}
\end{equation}
\end{subequations}
prohibiting backscattering. In
Eq.(\ref{sec:app_Coll_Int:eq:Coulomb_ME}),
$\omega=\ep_{\lambda,p}-\ep_{\lambda',p'}$ is the transfered energy
and $\bs{q}={\bs{p}}'-\bs{p}$ -- the transfered momentum.

Linearizing the collision integral~\eqref{colint} with respect to the
deviations \eqref{sec:hydro_micro:eq:delta_f} of the distribution
function from the local equilibrium
(\ref{sec:hydro_micro:eq:local_equilibrium}), we obtain the
operator\cite{Landau10,LinHydroBoris2015}
\begin{eqnarray}
&& 
\mathcal{C} \delta f_{\lambda,\bs{k}}
= \sum_{\nu,\lambda',\nu'}\int_{\bs{k},\bs{p}',\bs{k}'}
|M|^2 (2\pi)^3\delta(\ep_{p}+\ep_{k}-\ep_{p'}-\ep_{k'})
\delta(\bs{p}+\bs{k}-\bs{p}'-\bs{k}') 
\\
&&
\nonumber\\
&&
\qquad\qquad \qquad\qquad \qquad\qquad 
\times 
f^{(0)}_{\lambda,\bs{p}} f^{(0)}_{\nu,\bs{k}} 
\left[1-f^{(0)}_{\lambda',\bs{p}'}\right] 
\left[1-f^{(0)}_{\nu',\bs{k}'}\right]
\left[\delta f_{\lambda,\bs{p}}+\delta f_{\nu,\bs{k}}
-\delta f_{\lambda',\bs{p}'}-\delta f_{\nu',\bs{k}'}\right].
\nonumber
\label{sec:app_Coll_Int:eq:linear_coll_operator}
\end{eqnarray}

\subsection{Transport scattering times due to electron-electron interaction}
\label{sec:app_Gamma_1}

In this section we give explicit expressions for the scattering times
$\tau_{i}$ constituting the matrix collision integral in the space of
macroscopic currents $\bs{j}$ and $\bs{j}_I$, see
Eq.~\eqref{sec:hydro_macro_dissipative:eq:eom_currents_2}. These
equations are obtained by averaging the QKE with respect to $\bs{v}$
and $\lambda\bs{v}$. Therefore the right-hand side of
Eq.~\eqref{sec:hydro_macro_dissipative:eq:eom_currents_2} is given by
\begin{equation}
{\mathcal{C}}_J
\begin{pmatrix}
\delta\bs{j} \cr
\delta\bs{j}_I
\end{pmatrix}
=
\begin{pmatrix}
(\bs{v},\mathcal{C}\delta f^{(1)}) \cr
(\lambda\bs{v},\mathcal{C}\delta f^{(1)})
\end{pmatrix}.
\label{sec:app_Gamma_1:eq:coll_int_11}
\end{equation}
The scalar product $(\cdot,\cdot)$ was defined in
Eq.~\eqref{sec:hydro_macro:eq:scalar_product}.

Dissipative corrections to the macroscopic currents $\delta\bs{j}$ and
$\delta\bs{j}_I$ are determined by the non-equilibrium contribution to the
distribution function (\ref{sec:hydro_micro:eq:delta_f}) as
\begin{equation}
\delta\bs{j}_k = \big(\phi_k\bs{v},-T\delta f^{(1)}\partial_\ep f^{(0)}\big),
\label{sec:app_Gamma_1:eq:diss_current_def}
\end{equation}
such that 
\[
\delta\bs{j} = \delta\bs{j}_1, \qquad
\delta\bs{j}_I = \delta\bs{j}_2.
\]
Here we remind the reader that the terms proportional to $\bs{u}$ in
Eqs.~(\ref{j}) follow directly from the local equilibrium distribution
(\ref{sec:hydro_micro:eq:local_equilibrium}). The functions $\phi_k$ are
the modes \eqref{sec:hydro_micro:eq:modes}.

Now we can use the definition
(\ref{sec:app_Gamma_1:eq:diss_current_def}) to express the
coefficients $h^{(j)}$ in the non-equilibrium distribution
(\ref{sec:hydro_micro:eq:delta_f}) in terms of $\delta\bs{j}$ and
$\delta\bs{j}_I$. This allows us to find the explicit form of the
matrix collision integral ${\mathcal{C}}_J$, see
Eq.~(\ref{sec:hydro_macro_dissipative:eq:coll_integral_matrix}).
After some algebra, we find
\begin{equation}
[{\mathcal{C}}_J]_{lk}=\sum_{j=1}^{2}[\mathcal{M}^{-1}]_{jk}
(\phi_l{v}_\alpha,\mathcal{C}\phi_j{v}_\alpha),
\label{sec:app_Gamma_1:eq:coll_int_13}
\end{equation}
where the matrix $\mathcal{M}$ is given by
Eq.~(\ref{sec:hydro_macro_dissipative:eq:matrix_M}).

\begin{figure}[t]
\centerline{\includegraphics[width=0.8\linewidth]{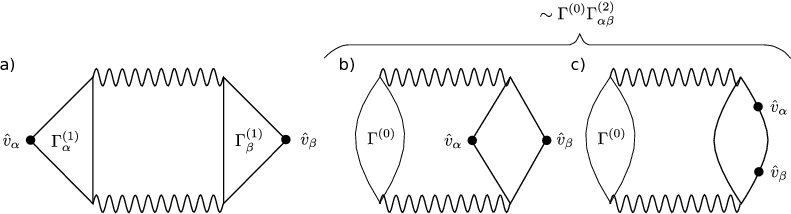}}
\caption{a) The Aslamazov-Larkin-type diagram corresponding to the
  term $\Gamma^{(1)}_\alpha\Gamma^{(1)}_\beta$. The product
  $\Gamma^{(0)}\Gamma^{(2)}_{\alpha\beta}$ comprises the
  Maki-Thompson-type diagram b) as well as self-energy correction c).}
\label{sec:app_Gamma_1:fig:gamma_fct_diag}
\end{figure}

The matrix elements in Eq.~(\ref{sec:app_Gamma_1:eq:coll_int_13}) can
be evaluated explicitly using the methods of
Refs.~\onlinecite{LinHydroBoris2015,SchuettDrag2013}. Noting that in
the integrated electron-electron collision integral the summation over
scattering states
$\{|\lambda,\bs{k}\rangle,|\lambda',\bs{k}'\rangle\}$ and
$\{|\nu,\bs{p}\rangle,|\nu',\bs{p}'\rangle\}$ separates,
we express the matrix elements as
\begin{equation}
(\phi v_\alpha,\mathcal{C} \phi'v_\beta) =
\frac{1}{16\pi}\int\!d\omega\int\! d^2 q\:
\frac{|V(\omega,q)|^2}{\sinh^2(\omega/2T)}
\left[
\Gamma^{(2)}_{\phi\phi',\alpha\beta}(\omega,q)\Gamma^{(0)}(\omega,q)
-\Gamma^{(1)}_{\phi,\alpha}(\omega,q)\Gamma^{(1)}_{\phi',\beta}(\omega,q)
\right].
\end{equation}
Here the vertex functions are defined as [$\lambda'=\text{sign}(\ep_{\lambda,p}+\omega)$],
\begin{subequations}
\label{gam}
\begin{equation}
\Gamma^{(0)}(\omega,q) = \frac{1}{T}\int_{\lambda,\bs{p}}\!
\delta(\ep_{\lambda,p}-\ep_{\lambda',\bs{p}+\bs{q}}+\omega)
\left(f^{(0)}_{\lambda,p}-f^{(0)}_{\lambda',\bs{p}+\bs{q}}\right)\Theta_{\lambda\bs{p};\lambda',\bs{p}+\bs{q}},
\label{sec:app_Gamma_1:eq:Gamma0} 
\end{equation}
\begin{equation}
\Gamma_{\phi,\alpha}^{(1)}(\omega,q) = \frac{1}{T}\int_{\lambda,\bs{p}}
\delta(\ep_{\lambda,p}-\ep_{\lambda',\bs{p}+\bs{q}}+\omega)
\left(f^{(0)}_{\lambda,p}-f^{(0)}_{\lambda',\bs{p}+\bs{q}}\right)\Theta_{\lambda,\bs{p};\lambda',\bs{p}+\bs{q}}
\big[\phi_{\lambda',\bs{p}+\bs{q}}\:\hat{v}_{\lambda',\bs{p}+\bs{q}}
-\phi_{\lambda\bs{p}}\:\hat{v}_{\lambda\bs{p}}\big]_\alpha,
\label{sec:app_Gamma_1:eq:Gamma1} 
\end{equation}
\begin{eqnarray}
&& 
\Gamma_{\phi\phi',\alpha\beta}^{(2)} = \frac{1}{T}\int_{\lambda,\bs{p}}
\delta(\ep_{\lambda,p}-\ep_{\lambda',\bs{p}+\bs{q}}+\omega)
\left(f^{(0)}_{\lambda,p}-f^{(0)}_{\lambda',\bs{p}+\bs{q}}\right)\Theta_{\lambda,\bs{p};\lambda',\bs{p}+\bs{q}} 
\\
&&
\nonumber\\
&&
\quad\quad\quad\quad\quad\quad 
\times
\big[\phi_{\lambda',\bs{p}+\bs{q}}\:\hat{v}_{\lambda',\bs{p}+\bs{q}}-\phi_{\lambda\bs{p}}\:\hat{v}_{\lambda\bs{p}}\big]_\alpha
\big[\phi'_{\lambda',\bs{p}+\bs{q}}\:\hat{v}_{\lambda',\bs{p}+\bs{q}}-\phi'_{\lambda\bs{p}}\:\hat{v}_{\lambda\bs{p}}\big]_\beta.
\label{sec:app_Gamma_1:eq:Gamma2}
\nonumber
\end{eqnarray}
\end{subequations}
The product $\Gamma^{(1)}_\alpha\Gamma^{(1)}_\beta$ can be represented
with the help of the Aslamazov-Larkin-type diagram in the Boltzmann
limit, whereas the product $\Gamma^{(0)}\Gamma^{(2)}_{\alpha\beta}$
contains the Maki-Thompson-type diagrams as well as self-energy
corrections, see Fig.~\ref{sec:app_Gamma_1:fig:gamma_fct_diag}.

The resulting values (\ref{taui}) are most conveniently calculated in
the local co-moving frame, where the hydrodynamic velocity entering
the local equilibrium distribution functions in Eqs.~(\ref{gam})
vanishes. The obtained results are then valid in arbitrary reference
frame based on the principle that the relaxation are independent
of the reference frame (generalizing the Galilean invariance to the
arbitrary spectrum).

\subsection{Dissipative corrections to the stress tensor} 
\label{sec:app_Gamma_2}

The collision integral ${\mathcal{C}}_\pi$ can be calculated along the
same lines as ${\mathcal{C}}_J$ in the previous Section. Averaging the
QKE with respect to the tensor quantities such as $v_\alpha v_\beta$,
we find the contribution of the collision integral in the form
similar to Eq.~(\ref{sec:app_Gamma_1:eq:coll_int_11})
\begin{equation}
{\mathcal{C}}_\pi
\begin{pmatrix}
\delta\Pi_{\alpha\beta} \cr
\delta\Pi_{I,\alpha\beta} \cr
T^{-1}\delta\Pi_{E,\alpha\beta}
\end{pmatrix}
	=
	\left(\begin{array}{c}
		({v}_\alpha{v}_\beta,\mathcal{C}\delta f^{(2)}) \\
		(\lambda{v}_\alpha{v}_\beta,\mathcal{C}\delta f^{(2)}) \\
		(\ep{v}_\alpha{v}_\beta/T,\mathcal{C}\delta f^{(2)})
	\end{array}\right).
	\label{sec:app_Gamma_2:eq:coll_int_21}
\end{equation}
The stress tensors were defined in
Eqs.~\eqref{sec:hydro_macro_dissipative:eq:def_stress_tensors} and
\eqref{sec:hydro_macro_dissipative:eq:def_stress_tensor_to_phi}.

Defining the deviations from equilibrium as
\begin{equation}
	\delta\Pi^{(k)}_{\alpha\beta}
	= \big(\phi_k {v}_\alpha {v}_\beta,-\delta f^{(2)}\partial_\ep f^{(0)}\big),
	\label{sec:app_Gamma_2:eq:stress_tensors_corrections}
\end{equation}
we can express the coefficients $g_{\alpha\beta}^{(j)}$ in the
non-equilibrium distribution function
(\ref{sec:hydro_micro:eq:delta_f}) in terms of
$\delta\Pi^{(k)}_{\alpha\beta}$. Similarly to the arguments presented
in the previous Section, this yields the explicit form of the matrix
collision integral ${\mathcal{C}}_\pi$:
\begin{equation}
	[{C}_\pi]_{lk}
	=
	2\sum_{j=1}^{3}
	[\mathcal{M}^{-1}]_{jk}
	(\phi_{l}I_{\alpha\beta},\mathcal{C} \phi_{j}I_{\alpha\beta}).
	\label{sec:app_Gamma_2:eq:coll_int_24}
\end{equation}
Here the matrix $\mathcal{M}$ is given by
Eq.~(\ref{sec:hydro_macro_dissipative:eq:matrix_M}) and the traceless
tensor $I_{\alpha\beta}$ is defined in
Eq.~(\ref{sec:hydro_macro_dissipative:eq:I}).

The matrix elements $(\phi_{l}I_{\alpha\beta},\mathcal{C}
\phi_{j}I_{\alpha\beta})$ can be evaluated similarly to
Eqs.~(\ref{gam}):
\begin{equation}
	(\phi I_{\alpha\beta},\mathcal{C} \phi'I_{\gamma\delta}) =
	\frac{1}{16\pi}\int\!d\omega\int\!d^2 q\:
	\frac{|V(\omega,q)|^2}{\sinh^2(\omega/2T)}
	\bigg[
		\Xi^{(2)}_{\phi\phi',\alpha\beta\gamma\delta}(\omega,q)\Gamma^{(0)}(\omega,q)
		-\Xi^{(1)}_{\phi,\alpha\beta}(\omega,q)\Xi^{(1)}_{\phi',\gamma\delta}(\omega,q)
	\bigg].
	\label{sec:app_Gamma_2:eq:ME_tensor_vertex_fct_1}
\end{equation}
Here the tensor vertex functions are [$\lambda'=\text{sign}(\ep_{\lambda,\bs{p}}-\omega)$],
\begin{subequations}
\label{xi}
\begin{equation}
\Xi_{\phi,\alpha\beta}^{(1)}(\omega,q)= \frac{1}{T}\int_{\lambda,\bs{p}}\!
\delta(\ep_{\lambda,p}-\ep_{\lambda',\bs{p}+\bs{q}}+\omega)
\left(f^{(0)}_{\lambda,p}-f^{(0)}_{\lambda',\bs{p}+\bs{q}}\right)\Theta_{\lambda,\bs{p};\lambda',\bs{p}+\bs{q}}
\big[\phi_{\lambda',\bs{p}+\bs{q}}\:I_{\alpha\beta,\bs{p}+\bs{q}}-\phi_{\lambda\bs{p}}\:I_{\alpha\beta,\bs{p}}\big],
\label{sec:app_Gamma_2:eq:Xi_1}
\end{equation}
\begin{eqnarray}
\label{sec:app_Gamma_2:eq:Xi_2}
&& 
\Xi_{\phi\phi',\alpha\beta\gamma\delta}^{(2)}(\omega,q) = \frac{1}{T}\int_{\lambda,\bs{p}}\!
\delta(\ep_{\lambda,p}-\ep_{\lambda',\bs{p}+\bs{q}}+\omega)
\left(f^{(0)}_{\lambda,p}-f^{(0)}_{\lambda',\bs{p}+\bs{q}}\right)\Theta_{\lambda,\bs{p};\lambda',\bs{p}+\bs{q}}
\\
&&
\nonumber\\
&& 
\qquad\qquad\qquad\qquad
\times
\big[\phi_{\lambda',\bs{p}+\bs{q}}\:I_{\alpha\beta,\bs{p}+\bs{q}}-\phi_{\lambda\bs{p}}\:I_{\alpha\beta,\bs{p}}\big]
\big[\phi'_{\lambda',\bs{p}+\bs{q}}\:I_{\gamma\delta,\bs{p}+\bs{q}}-\phi'_{\lambda\bs{p}}\:I_{\gamma\delta,\bs{p}}\big].
\nonumber
\end{eqnarray}
\end{subequations}
For further calculations it is useful to express the tensor $I_{\alpha\beta}$
in terms of the basis vectors $\{\hat{q}=\bs{q}/q,\hat{q}_\perp=\hat{z}\times\hat{q}\}$,
\begin{equation}
	I_{\alpha\beta}
	= A_{\bs{k},\bs{q}} (2\hat{q}_\alpha\hat{q}_\beta-\delta_{\alpha\beta})
	+ B_{\bs{k},\bs{q}}(\hat{q}_{\perp,\alpha}\hat{q}_{\beta}+\hat{q}_{\alpha}\hat{q}_{\perp,\beta})
	\: ,
	\label{sec:app_Gamma_2:eq:tensor_rep}
\end{equation}
where
\begin{equation}
A_{\bs{k},\bs{q}} = \left(\frac{(\bs{k}\cdot\bs{q})^2}{(kq)^2} - 1 \right) + \frac{1}{2}
	= \tilde{A}_{\bs{k},\bs{q}} + \frac{1}{2} ,
\qquad
B_{\bs{k},\bs{q}} = \frac{(\bs{k}\cdot\bs{q}_\perp)(\bs{k}\cdot\bs{q})}{k^2q^2}.
	\label{sec:app_Gamma_2:eq:B_def}
\end{equation}
Due to the conservation laws of the electron-electron interaction we
effectively have $A\rightarrow\tilde{A}$.  Using the $\delta$-function
in Eqs.~\eqref{sec:app_Gamma_2:eq:Xi_1} and
\eqref{sec:app_Gamma_2:eq:Xi_2} one obtains ($\ep=\ep_{\lambda,k}$),
\begin{equation}
	\tilde{A}_{\bs{k},\bs{q}} =
	(\omega^2-q^2)\:\frac{(2\ep+\omega)^2-q^2}{8 \ep^2 q^2}
	\: .
	\label{sec:app_Gamma_2:eq:A_expl}
\end{equation}
Furthermore, the coefficient $B$ drops out in the vertex function
$\Xi^{(1)}$ since it is antisymmetric in the angle between $\bs{q}$
and $\bs{k}$.  In the tensor vertex function
$\Xi^{(2)}_{\alpha\beta\gamma\delta}(\omega,q)$ we get a separate
contribution from $A$ and $B$ but they are orthogonal.  For $B$ we
obtain with the help of the $\delta$-functions
($\ep=\ep_{\lambda,k}$),
\begin{equation}
B_{\bs{k},\bs{q}} = \text{sign}(\bs{k}\cdot\hat{q}_\perp)
\frac{\sqrt{\left(q^2-\omega^2\right)[(2\ep+\omega)^2-q^2]}\:(\omega^2-q^2-2\ep\omega)}{4\ep^2 q^2}.
	\label{sec:app_Gamma_2:eq:B_expl}
\end{equation}
Finally, with the help of the angular averages
\[
\int\!d\varphi_q\:\hat{q}_\alpha\hat{q}_\beta\hat{q}_\gamma \hat{q}_\delta
	= \frac{\pi}{4}(\delta_{\alpha\gamma}\delta_{\beta\delta}+\delta_{\alpha\delta}\delta_{\beta\gamma}+\delta_{\alpha\beta}\delta_{\gamma\delta}),
\]
\[ 
\int\!d\varphi_q\:(2\hat{q}_\alpha\hat{q}_\beta-\delta_{\alpha\beta})(2\hat{q}_\gamma\hat{q}_\delta-\delta_{\gamma\delta})
	= \pi(\delta_{\alpha\gamma}\delta_{\beta\delta}+\delta_{\alpha\delta}\delta_{\beta\gamma}-\delta_{\alpha\beta}\delta_{\gamma\delta}),
\]
\[ 
\int\!d\varphi_q\:(\hat{q}_{\perp,\alpha}\hat{q}_{\beta}+\hat{q}_{\alpha}\hat{q}_{\perp,\beta})(\hat{q}_{\perp,\gamma}\hat{q}_{\delta}+\hat{q}_{\gamma}\hat{q}_{\perp,\delta})
	= \pi(\delta_{\alpha\gamma}\delta_{\beta\delta}+\delta_{\alpha\delta}\delta_{\beta\gamma}-\delta_{\alpha\beta}\delta_{\gamma\delta})
\]
and the projected quantities $\Xi^{(1,2)}$ obtained after averaging
Eqs.~(\ref{xi}) over the angle $\varphi_q$ of the transfered momentum
$\bs{q}$,
\begin{subequations}
\begin{equation}
\Xi_{\phi}^{(1)}(\omega,q)= \frac{1}{T}\int_{\lambda,\bs{p}}\!
\delta(\ep_{\lambda,p}-\ep_{\lambda',\bs{p}+\bs{q}}+\omega)
\left(f^{(0)}_{\lambda,p}-f^{(0)}_{\lambda',\bs{p}+\bs{q}}\right)\:\Theta_{\lambda,\bs{p};\lambda',\bs{p}+\bs{q}}
\big[\phi_{\lambda',\bs{p}+\bs{q}}\:\tilde{A}_{\bs{k}+\bs{q},\bs{q}}-\phi_{\lambda\bs{p}}
\tilde{A}_{\bs{k},\bs{q}}\big],
\label{sec:app_Gamma_2:eq:Xi_1_parallel} 
\end{equation}
\begin{eqnarray}
\label{sec:app_Gamma_2:eq:Xi_2_parallel}&& 
\Xi_{\parallel,\phi\phi'}^{(2)}(\omega,q) = \frac{1}{T}\int_{\lambda,\bs{p}}\!
\delta(\ep_{\lambda,p}-\ep_{\lambda',\bs{p}+\bs{q}}+\omega)
\left(f^{(0)}_{\lambda,p}-f^{(0)}_{\lambda',\bs{p}+\bs{q}}\right)\Theta_{\lambda,\bs{p};\lambda',\bs{p}+\bs{q}} 
\\
&&
\nonumber\\
&&
\qquad\qquad\quad\quad\quad\quad 
\times
\big[\phi_{\lambda',\bs{p}+\bs{q}}\:\tilde{A}_{\bs{p}+\bs{q},\bs{q}}
-\phi_{\lambda\bs{p}}\:\tilde{A}_{\bs{p},\bs{q}}\big]
\big[\phi'_{\lambda',\bs{p}+\bs{q}}\:\tilde{A}_{\bs{p}+\bs{q},\bs{q}}
-\phi'_{\lambda\bs{p}}\:\tilde{A}_{\bs{p},\bs{q}}\big],
\nonumber
\end{eqnarray}
\begin{eqnarray}
\label{sec:app_Gamma_2:eq:Xi_2_perp}
&& 
\Xi_{\perp,\phi\phi'}^{(2)}(\omega,q) = \frac{1}{T}\int_{\lambda,\bs{p}}\!
\delta(\ep_{\lambda,p}-\ep_{\lambda',\bs{p}+\bs{q}}+\omega)
\left(f^{(0)}_{\lambda,p}-f^{(0)}_{\lambda',\bs{p}+\bs{q}}\right)\Theta_{\lambda,\bs{p};\lambda',\bs{p}+\bs{q}}
\\
&&
\nonumber\\
&&
\qquad\qquad\quad\quad\quad\quad \times
\big[\phi_{\lambda',\bs{p}+\bs{q}}\:B_{\bs{p}+\bs{q},\bs{q}}-\phi_{\lambda\bs{p}}\:B_{\bs{p},\bs{q}}\big]
\big[\phi'_{\lambda',\bs{p}+\bs{q}}\:B_{\bs{p}+\bs{q},\bs{q}}-\phi'_{\lambda\bs{p}}\:B_{\bs{p},\bs{q}}\big],
\nonumber
\end{eqnarray}
\end{subequations}
we can write the matrix elements as
\begin{eqnarray}
\label{sec:app_Gamma_2:eq:ME_tensor_vertex_fct_2}
&&
(\phi I_{\alpha\beta},\mathcal{C} \phi'I_{\gamma\delta}) =\frac{1}{16\pi}
(\delta_{\alpha\gamma}\delta_{\beta\delta}+\delta_{\alpha\delta}\delta_{\beta\gamma}
-\delta_{\alpha\beta}\delta_{\gamma\delta})
\int\!d\omega\int\!d^2 q
\frac{|V(\omega,q)|^2}{\sinh^2(\omega/2T)} 
\\
&&
\nonumber\\
&&
\qquad\qquad
\bigg[
\Xi^{(2)}_{\parallel,\phi\phi'}(\omega,q)\Gamma^{(0)}(\omega,q)
+\Xi^{(2)}_{\perp,\phi\phi'}(\omega,q)\Gamma^{(0)}(\omega,q)
-\Xi^{(1)}_{\phi}(\omega,q)\Xi^{(1)}_{\phi'}(\omega,q)
\bigg].
\nonumber
\end{eqnarray}
Here we can drop the terms proportional to $\delta_{\alpha\beta}$
since the energy stress tensors are traceless.  Due to their symmetry
in $\alpha\leftrightarrow\beta$ we effectively have
\begin{equation}
\delta_{\alpha\gamma}\delta_{\beta\delta}+\delta_{\alpha\delta}\delta_{\beta\gamma}
-\delta_{\alpha\beta}\delta_{\gamma\delta}
\rightarrow
2\delta_{\alpha\gamma}\delta_{\beta\delta}.
\end{equation}
The matrix elements (\ref{sec:app_Gamma_2:eq:ME_tensor_vertex_fct_2})
determine the quantities ${\cal C}_{\pi,ij}$, Eqs.~(\ref{dpidp}),
which in turn determine the viscosity
(\ref{sec:hydro_macro_dissipative:eq:viscosity}).

\section{Linear response functions}
\label{sec:app_lin_response}

In linear response we linearize the hydrodynamic equations with
respect to the linear fluctuations of the hydrodynamic quantities,
$\delta n$, $\delta n_I$, $\delta n_E$, $\delta u$:
\begin{equation}
n \rightarrow n + \delta n, \qquad
n_I \rightarrow n_I + \delta n_I, \qquad
n_E \rightarrow n_E + \delta n_E, \qquad
u \rightarrow \delta u.
\end{equation}
We furthermore introduce the response functions to the external
perturbation $\bs{E}=-i\bs{q}\varphi$
\begin{equation}
\delta n = \chi_n \varphi,
\qquad
\delta n_I = \chi_I \varphi,
\qquad
\delta n_E = T \chi_E \varphi,
\qquad
\delta u = -i\frac{\bs{q}T}{qn_E} \chi_u \varphi.
\end{equation}

Linearizing the continuity equations (\ref{cont2}) and the Navier-Stokes
equation (\ref{sec:Euler:eq:Euler_eq_2}), we find the matrix equation 
for the response functions $\chi_i$:
\begin{equation}
\begin{pmatrix}
-i\omega + \frac{\tau_1}{2}q^2 - 2\pi eq \sigma_0 & \frac{\tau_2}{2}q^2 & 
    -\left(\frac{n \tau_1+n_I \tau_2}{3n_E}T\right) q^2 & \frac{nT}{n_E}q \cr
\frac{\tau_3}{2} q^2 - 2\pi eq \sigma_0^* & -i\omega + \frac{\tau_4}{2}q^2 & 
    -\left(\frac{n_I\tau_4+n\tau_3}{3n_E}T\right)q^2 & \frac{n_IT}{n_E}q  \cr
0 & 0 & -i\omega & \frac{3}{2} q \cr
-\frac{4\pi e^3 n}{3T}q & 0 & -\frac{q}{3} & -i\omega +\tau_{\rm dis}^{-1} + \frac{2\eta}{3n_E} q^2
\end{pmatrix}
	\left(
	\begin{array}{c}
		\chi_n \\
		\chi_I \\
		\chi_E \\
		\chi_u
	\end{array}
	\right)
	=
	\left(
	\begin{array}{c}
		-q^2\sigma_0/e \\
		-q^2\sigma_0^*/e \\
		0 \\
		\frac{2e n}{3T} q
	\end{array}
	\right),
	\label{sec:app_lin_response:eq:MatrixEquation}
\end{equation}
where [cf. Eq.~(\ref{sec:lin_resp:eq:quantum_cond})]
\[
\sigma_0=e^2\!\left[\tau_1\left(\frac{\partial_\mu n}{2}-\frac{2n^2}{3n_E}\right)
		\!+\!\tau_2\left(\frac{\partial_\mu n_I}{2}-\frac{2n n_I}{3n_E}\right)\right], 
\qquad
\sigma_0^*=e^2\!\left[\tau_3\left(\frac{\partial_\mu n}{2}-\frac{2n^2}{3n_E}\right)
		\!+\!\tau_4\left(\frac{\partial_\mu n_I}{2}-\frac{2n n_I}{3n_E}\right)\right].
\]

The dispersion (\ref{sec:lin_resp:eq:disp_rel}) of the collective
modes follows from zeros of the determinant of the matrix in the
left-hand side of Eq.~(\ref{sec:app_lin_response:eq:MatrixEquation}). 

In contrast to the energy waves and plasmons, which describe the
response of the system to an external perturbation, the conductivity
of an infinite system is defined as the response to the total electric
field. Consequently, in order to find the conductivity
(\ref{sec:lin_resp:eq:lin_resp_cond}), we need to consider the
irreducible response functions, which satisfy the equation similar to
Eq.~(\ref{sec:app_lin_response:eq:MatrixEquation}), but without the
Vlasov terms in the left column of the matrix in the left hand side.
Then the conductivity is found from the Ohm's law
\begin{equation}
\delta\bs{j} = (-i\bs{q}\varphi)
\left[\sigma_0/e + \frac{1}{2}(\tau_1\chi_n+\tau_2\chi_I)-
\frac{n \tau_1+n_I \tau_2}{3n_E}T \chi_E
+ \frac{nT}{qn_E}\chi_n\right],
\end{equation}
where $\varphi$ is now the total potential in the system (including the
self-consistent Vlasov contribution).

\twocolumngrid

\bibliography{p_nonlinear_hydro}

\end{document}